\documentclass[usegraphicx]{mn2e} 
\usepackage{lscape}
\usepackage{subfigure}  
\usepackage{longtable}  
\usepackage{graphicx}  
\usepackage{graphics}  
\usepackage{amssymb} 
\usepackage{keyval}  
\usepackage{trig}  
\def\beq{\begin{equation}}  
\def\eeq{\end{equation}}  
\def\bey{\begin{eqnarray}}  
\def\eey{\end{eqnarray}}

\def\kms{\, {\rm km \, s}^{-1} }

\def\a0{$a_0$}

\def\topmargin{-0.9cm}

\title[Chemically tagging the Hyades stream]{Chemically tagging the Hyades stream: does it partly originate from the Hyades cluster?\thanks{Based on observations obtained with the HERMES/Mercator
spectrograph/telescope installed at the Roque de los Muchachos Observatory, La Palma, Spain} }  
\author[L. Pomp\'eia et al.]    {L. Pomp\'eia$^{1,2}$, T. Masseron$^{1}$, 
B. Famaey$^{3,4}$,  S. Van Eck$^{1}$,  A. Jorissen$^{1}$, I. Minchev$^{5}$, \newauthor A. Siebert$^{3}$, C. Sneden$^{6}$,  J.R.D. L\'epine$^{7}$,  C. Siopis$^{1}$, G. Gentile$^{1,8}$, T. Dermine$^{1}$, \newauthor  E. Pasquato$^{1}$, H. Van Winckel$^{9}$, C. Waelkens$^{9}$, G. Raskin$^{9}$, S. Prins$^{9}$, W. Pessemier$^{9}$, \newauthor H. Hensberge$^{10}$, Y. Fr\'emat$^{10}$, L. Dumortier$^{10}$, O. Bienaym\'e$^{3}$\\  \\
$^{1}$Institut d'Astronomie et d'Astrophysique, Universit\'e Libre  
de Bruxelles, CP 226, Bruxelles, Belgium\\
$^{2}$Universidade do Vale do Para\'\i ba, Av. Shishima Hifumi, 2911, S\~ao, Jos\'e dos Campos, 
  12244-000 SP, Brazil\\
$^{3}$Observatoire Astronomique, Universit\'e de Strasbourg, CNRS UMR 7550, 67000 Strasbourg, 
   France \\
$^{4}$AIfA, Universt\"at Bonn, 53121 Bonn, Germany\\
$^{5}$AIP, An der Sterwarte 16, 14482 Potsdam, Germany\\
$^{6}$Department of Astronomy and McDonald Observatory, The University of Texas, Austin, 
   TX 78712, USA \\
$^{7}$Universidade de S\~ao Paulo, IAG, C.P. 3386, 01060-970 S\~ao Paulo,
Brazil\\
$^{8}$Sterrenkundig Observatorium, Universiteit Gent, Krijgslaan 281, 9000 Gent, Belgium\\
$^{9}$Instituut voor Sterrekunde, Katholiek Universiteit Leuven,
Celestijnenlaan 200D, 3001 Leuven, Belgium\\
$^{10}$Royal Observatory of Belgium, Avenue Circulaire 3, 1180 Bruxelles, Belgium\\} 
  
\begin{document}  
  
\date{Accepted ... Received ... ; in original form ...}  
  
\pagerange{\pageref{firstpage}--\pageref{lastpage}} \pubyear{2011}  
  
\maketitle  
  
\label{firstpage}  
  
\begin{abstract}  
The Hyades stream has long been thought to be a dispersed vestige of the Hyades cluster. However, recent analyses of the parallax distribution, of the mass function, and of the action-space distribution of stream stars have shown it to be rather composed of orbits trapped at a resonance of a density disturbance. This resonant scenario should leave a clearly different signature in the element abundances of stream stars than the dispersed cluster scenario, since the Hyades cluster is chemically homogeneous. Here, we study the metallicity as well as the element abundances of Li, Na, Mg, Fe, Zr, Ba, La, Ce, Nd, and Eu for a random sample of stars belonging to the Hyades stream, and compare them with those of stars from the Hyades cluster. From this analysis: (i) we independently confirm that the Hyades stream cannot be solely composed of stars originating in the Hyades cluster; (ii) we show that {\it some} stars (namely 2/21) from the Hyades stream nevertheless have abundances compatible with an origin in the cluster; (iii) we emphasize that the use of Li as a chemical tag of the cluster origin of main-sequence stars is very efficient in the range $5500 \, {\rm K} \le T_{\rm eff} \le 6200 \, {\rm K}$, since the Li sequence in the Hyades cluster is very tight, while at the same time spanning a large abundance range; (iv) we show that, while this evaporated population has a metallicity excess of $\sim 0.2$~dex w.r.t. the local thin disk population, identical to that of the Hyades cluster, the remainder of the Hyades stream population has still a metallicity excess of $\sim 0.06$ to 0.15~dex, consistent with an origin in the inner Galaxy; (v) we show that the Hyades stream can be interpreted as an inner 4:1 resonance of the spiral pattern: this then also reproduces an orbital family compatible with the Sirius stream, and places the origin of the Hyades stream up to 1~kpc inwards from the solar radius, which might explain the observed metallicity excess of the stream population.
\end{abstract}  
  
\begin{keywords}  
Galaxy: evolution - clusters and associations - disk - solar neighbourhood  
\end{keywords}  
  
\section{Introduction}  
\label{sec:intro}

It has been known for a long time that a spatially unbound group of
stars in the solar neighbourhood is sharing the same kinematics as the
Hyades open cluster (e.g., Eggen 1958). Assuming that it is a vestige
of an initially more massive Hyades cluster which dispersed with time, with the result that its distribution function is still presently evolving towards equilibrium and does not yet satisfy the Jeans theorem, Eggen called this kinematically cold group the Hyades supercluster. More generally, it is called the Hyades moving group or Hyades stream, since Eggen's hypothesis that kinematic groups of this type are indeed such cluster remnants has been largely debated for many years. Actually, such groups may also be generated by a number of global dynamical mechanisms. Most of the mechanisms able to generate such unbound groups of stars moving
in a peculiar fashion are linked with the non-axisymmetry of the Galaxy,
namely with the presence of a rotating central bar (e.g., Dehnen 1998,
Fux 2001, Minchev et al. 2010) and of spiral arms (e.g., Quillen \&
Minchev 2005, Antoja et al. 2009), or both (see Quillen 2003, Minchev
\& Famaey 2010). 

A way to discriminate between these two hypotheses is the ``chemical tagging"  of stars belonging to the moving groups (Freeman \& Bland-Hawthorn 2002, De Silva et al. 2009). With this technique, a true Eggen moving group (or true supercluster) was recently tracked down, namely the HR1614 moving group (Eggen 1978, Feltzing \& Holmberg 2000), in which the high level of chemical homogeneity is supporting the case that it is a relic of an ancient star-forming event (De Silva et al. 2007). Conversely, the Hercules stream was found to be composed of stars whose abundance pattern match that of the disk stars (Bensby et al. 2007), which could be interpreted as being associated with the outer Lindblad resonance of the bar (Dehnen 1998, Fux 2001). 

Here, we intend to analyze the abundance trends of a sample of stars belonging to the Hyades stream, in order to check whether it is compatible with Eggen's scenario, with the resonant scenario, or with a mix of both. Indeed, the evaporation of open clusters and the dynamical
perturbations linked with density perturbers are not mutually incompatible phenomena. Stars originating from a 600 Myr old cluster had time to disperse over more than 500~pc in the disk (e.g., Bland-Hawthorn et al. 2010), but have retained the same guiding radius and same tangential velocity $V$ (Woolley 1961). This does not prevent an overdensity in velocity space due to orbits trapped at resonance to overlap with this cluster remnant.

\begin{figure}
\includegraphics[width=8cm]{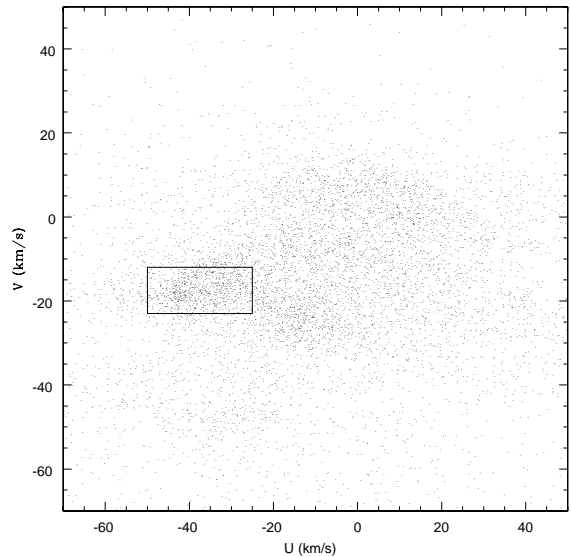}
\caption{
\label{Fig:GC}
The full Geneva-Copenhagen (GC) sample from Holmberg et al. (2009) in the $UV$ plane. The Hyades box is also indicated at $-23 \, \kms$ $\leq V \leq -12 \, \kms$ and $-50 \, \kms \, \leq U \leq -25 \, \kms$.
}
\end{figure}

The study of the
kinematics of K and M giant stars in the solar neighbourhood, combining
CORAVEL radial velocities with Hipparcos parallaxes and proper motions
(Famaey et al. 2005), already indicated that the Hyades stream
could not be {\it solely} composed of coeval stars evaporated from the
primordial Hyades cluster, because the Hertzsprung-Russell diagram was
very similar for stars in the stream and in the field, indicating
a wide range of ages for stars belonging to the stream. This was
later confirmed by a detailed analysis of the distribution of stars in
parallax space as compared to the parallax corresponding to the Hyades
cluster's 600~Myr isochrone (Famaey, Siebert \& Jorissen 2008). Bovy \& Hogg~(2010) recently reached the same conclusion. Moreover, at about the same time, radial velocities, masses, and
metallicities of more than 14000 F and G dwarfs stars were also
published as the {\it Geneva-Copenhagen} (GC) {\it Survey} (Nordstr\"om et al.
2004, Holmberg et al. 2007, 2009), supplementing the Hipparcos parallaxes and proper
motions. The distribution of these main-sequence
stars in velocity space confirmed that the Hyades stream was the most prominent feature on top of the background velocity ellipsoid (Fig.~\ref{Fig:GC}). Sellwood~(2010) then estimated the actions of individual stars from their observed coordinates and velocities, and examined the stellar distribution in the $J_R$-$J_\phi$ action space. He found about 5\% of the GC stars concentrated along a resonance line in action space, which was interpreted as a signature of scattering at the inner Lindblad resonance of a spiral pattern\footnote{Most likely a 4:1 inner resonance, in order to prevent the corotation from being too far out in the disk.}. This overdensity in action space precisely corresponds to the Hyades stream in velocity space, in accordance with the model of Quillen \& Minchev~(2005). What is more, Famaey et al.~(2007) compared the mass distribution of stars belonging to the Hyades stream with the initial mass function of the Hyades cluster and with the present-day mass function of
the galactic disk, showing that it was in disagreement with the
former and in agreement with the latter, thus also favoring a dynamical,
resonant origin for the stream. However, the latter analysis was compatible with a proportion of at least 15\% of stream 
stars actually being past members of the Hyades cluster. It was also found that, while thin disk stars have a mean metallicity [Fe/H]$\simeq -0.15$, the mean metallicity of stars moving with the Hyades (among which only $\sim$25\% of field stars from the background velocity ellipsoid, the rest constituting the Hyades overdensity in velocity space) was [Fe/H]~$=-0.06$. Since the Hyades cluster is also more metal rich than the field, with a mean [Fe/H]~$=+0.14$ (Cayrel de Strobel et al. 1997, Perryman et al. 1998, Grenon 2000), this could also argue in favor of a large proportion of stars in the Hyades stream originating from the Hyades cluster. However, this higher metallicity of the Hyades stream could also indicate that its stars originate from the inner Galaxy.

In this paper, we aim at determining the chemical abundances of stars belonging to the Hyades stream, in order to yield constraints on their origin (from the cluster or from the field of the disk, and in the latter case, from which typical galactocentric radius). In this way we shall be able to {\it independently} confirm from chemical tagging that the Hyades stream cannot be solely composed of stars originating in the Hyades cluster, and we shall be able to provide the first {\it direct} evidence for two distinct populations inside the stream. We also aim at making a preliminary analysis of the respective characteristics of these two populations. In Sect.~2, we present the sample of stars that we are going to analyze, and in Sect.~3 we briefly describe the observations performed with the HERMES/Mercator spectrograph (Raskin et al. 2011). Stellar parameters and abundances are respectively determined in Sects.~4 and 5. We then investigate whether stars in the stream are compatible with being evaporated from the Hyades cluster in Sects.~6 and 7, and discuss these results and their consequences in Sect.~8. Conclusions are drawn in Sect.~9.

\begin{table*}
\caption[]{\label{Tab:sample}
Data for the analyzed stellar sample.
}
\begin{tabular}{lrcccrlccrcccccr}
\hline
\hline
HD &    \multicolumn{1}{c}{$d$} & $M_V$ & $U$ & $V$ & \multicolumn{1}{c}{$W$} & $m_{v}$ & $b-y$ & $\cal{M}$ & \multicolumn{1}{c}{$V_{\rm rot} \sin i$}\\
     &   \multicolumn{1}{c}{(pc)} &          & (km/s) & (km/s) & \multicolumn{1}{c}{(km/s)} & (mag) & & ($\cal{M}_\odot$)
     & \multicolumn{1}{c}{(km/s)}\\
\hline\\
&&&\multicolumn{5}{c}{Sure Hyades cluster members}\\
\hline\\
18632    &      23 &  6.13 &   -42    & -19& -1      &  7.978 &        &
 0.87 & 3\\
19902    &      40 &  5.19 &   -41 &   -19 &	-1 &  8.171 &  0.452 &    0.89 &      2 \\   
26756    &      46 &  5.13 &   -41 &   -18 &	-3 &  8.457 &  0.431 &    0.93 &      5 \\  
26767    &      43 &  4.88 &   -41 &   -18 &	-3 &  8.045 &  0.405 &    0.98 &      5 \\  
HIP 13806&   39 &  5.94 &   -41 &   -17 &   -2  &  8.90  &        &
 0.89 & 4\\
\hline\\
&&&\multicolumn{5}{c}{Possible Hyades cluster members}\\
\hline\\
20430    &      46 &  4.07 &   -42 &   -23 &	-2 &  7.386 &  0.364 &    1.10 &      6 \\  
20439    &      42 &  4.66 &   -41 &   -21 &	-4 &  7.766 &  0.395 &    1.17 &      6 \\ 
26257    &      58 &  3.84 &   -35 &   -13 &	-5 &  7.639 &  0.345 &    1.25 &      8 \\   
HIP 13600&      53 &  5.21 &  -40  & -13  &  -2    &  8.83  &        &
 1.00 & 2\\
\hline\\
&&&\multicolumn{5}{c}{Hyades velocity box}\\
\hline\\
25680     &      17 &  4.79 &   -25 &   -14 &    -6 &  5.903 &  0.399 &    0.96 &      3  \\
42132     &     111 &  1.52 &   -43 &   -16 &   -23 &  6.675 &  0.525 &    0.89 &      3  \\
67827     &      45 &  3.39 &   -35 &  	-14 &	-9  &  6.580 &	0.368 &	   1.34 &	5 \\
86165    &      51 &  4.39 &   -35 &   -17 &   -16 &  7.926 &  0.386 &    0.97 &	2 \\  
89793    &      63 &  4.87 &   -33 &   -20 &    12 &  8.853 &  0.424 &    0.96 &	1 \\	
90936    &      56 &  4.62 &   -42 &   -18 &   -14 &  8.371 &  0.389 &    0.96 &	3 \\	
103891    &      58 &	2.79 &	-28 &	-17 &	7 &	6.591 &	0.356 &	   1.35 &	4  \\
108351    &      88 &	3.19 &	-50 &	-19 &	5 &	7.905 &	0.327 &	   1.40 &	9  \\
133430   &      57 &  4.79 &   -30 &   -16 &    -5 &  8.559 &  0.413 &    0.90 &	2 \\	
134694   &     134 &  2.87 &   -43 &   -20 &   -17 &  8.502 &  0.336 &    1.34 &    10 \\   
142072   &      39 &  4.90 &   -29 &   -15 &    -1 &  7.854 &  0.417 &    0.90 &	7 \\	
149028   &      48 &  5.10 &   -37 &   -16 &    -1 &  8.530 &  0.454 &    0.91 &	2 \\  
149285   &     196 &  2.36 &   -34 &   -14 &   -14 &  8.825 &  0.337 &    1.32 &	8 \\   
151766   &     109 &  2.42 &   -35 &   -13 &   -20 &  7.606 &  0.348 &    1.46 &	8 \\       
155968   &      55 &  4.72 &   -30 &   -14 &     2 &  8412  &  0.423 &    0.96 &	3 \\	
157347   &      20 &  4.83 &   -28 &   -17 &   -20 &  6.287 &  0.425 &    0.90 &	1 \\   
162808   &      63 &  4.41 &   -31 &   -12 &    -2 &  8.423 &  0.393 &    1.00 &	6 \\	
171067   &      26 &  5.13 &   -46 &   -16 &   -14 &  7.205 &  0.424 &    0.90 &	1 \\	
180712   &      45 &  4.73 &   -25 &   -12 &    -6 &  7.977 &  0.395 &    0.88 &	3 \\   
187237   &      26 &  4.78 &   -36 &   -20 &    14 &  6.877 &  0.409 &    0.90 &	2 \\  
189087   &      27 &  5.75 &   -42 &   -15 &     5 &  7.886 &  0.483 &    0.81 &	2 \\  
\hline
\end{tabular} 
\end{table*}

\section{Sample}
\label{Sect:sample}

\begin{figure}
\includegraphics[width=8cm]{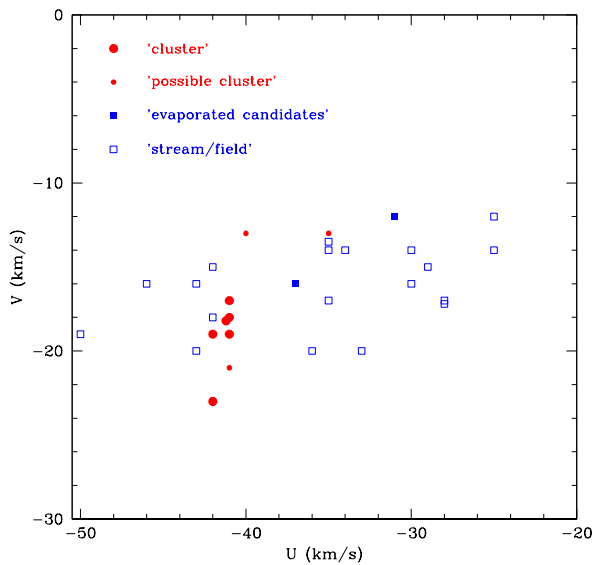}
\caption{\label{Fig:UV}
The present stellar sample in the $U-V$ plane.  The meaning of the
symbols is
given  in the figure (see Sect.~\ref{Sect:evap} for more details).
}
\end{figure}

Fig.~\ref{Fig:GC} presents the distribution
of all stars from the GC survey in the $UV$-plane ($U$ is the velocity
towards the galactic centre, $V$ the velocity in the direction of
Galactic rotation, both with respect to the Sun), and the Hyades
overdensity corresponds to the box defined by $-23 \, \kms \, \leq V \leq
-12 \, \kms$ and $-50 \, \kms \, \leq U \leq -25 \, \kms$ (see Famaey et al. 2007): this box corresponds to a region in which the density 
of the GC survey in $UV$ velocity space is higher than $\sim 4$ stars/(km/s)$^2$.

A randomly chosen stellar sample within this velocity box has been selected among the GC survey stars with $b-y \ge 0.3$ (or equivalently, $B-V \ge 0.5$, according to Tables~15.7 and 15.10 of Drilling \& Landolt 1999) and $V_{\rm rot} \sin i \le 10$~$\kms$, the latter condition in order to ease the abundance determination. Members of the Hyades cluster, taken from the De Silva et al. (2006) sample 
have been added to this GC kinematical sample, in order to compare the element abundances in the stream and in the cluster. The criterion on the colour index of the stream stars is necessary to  avoid any bias from the rejection of the fast rotators from the stream: since Paulson et al. (2003; their Fig.~5) have shown that in the Hyades cluster, all stars (but one) with $B - V \ge 0.5$ have $V_{\rm rot} \sin i \le 10$~$\kms$, so that few if any stars evaporated from the cluster have been left aside by selecting only slow rotators with $B - V \ge 0.5$ in the stream.

Among the Hyades cluster members we have selected, 4 are controversial (see Table~\ref{Tab_param}): they are spatially associated with the cluster but they might not belong to it based on kinematics (de Bruijne et al. 2001).

\begin{figure}
\includegraphics[width=8cm]{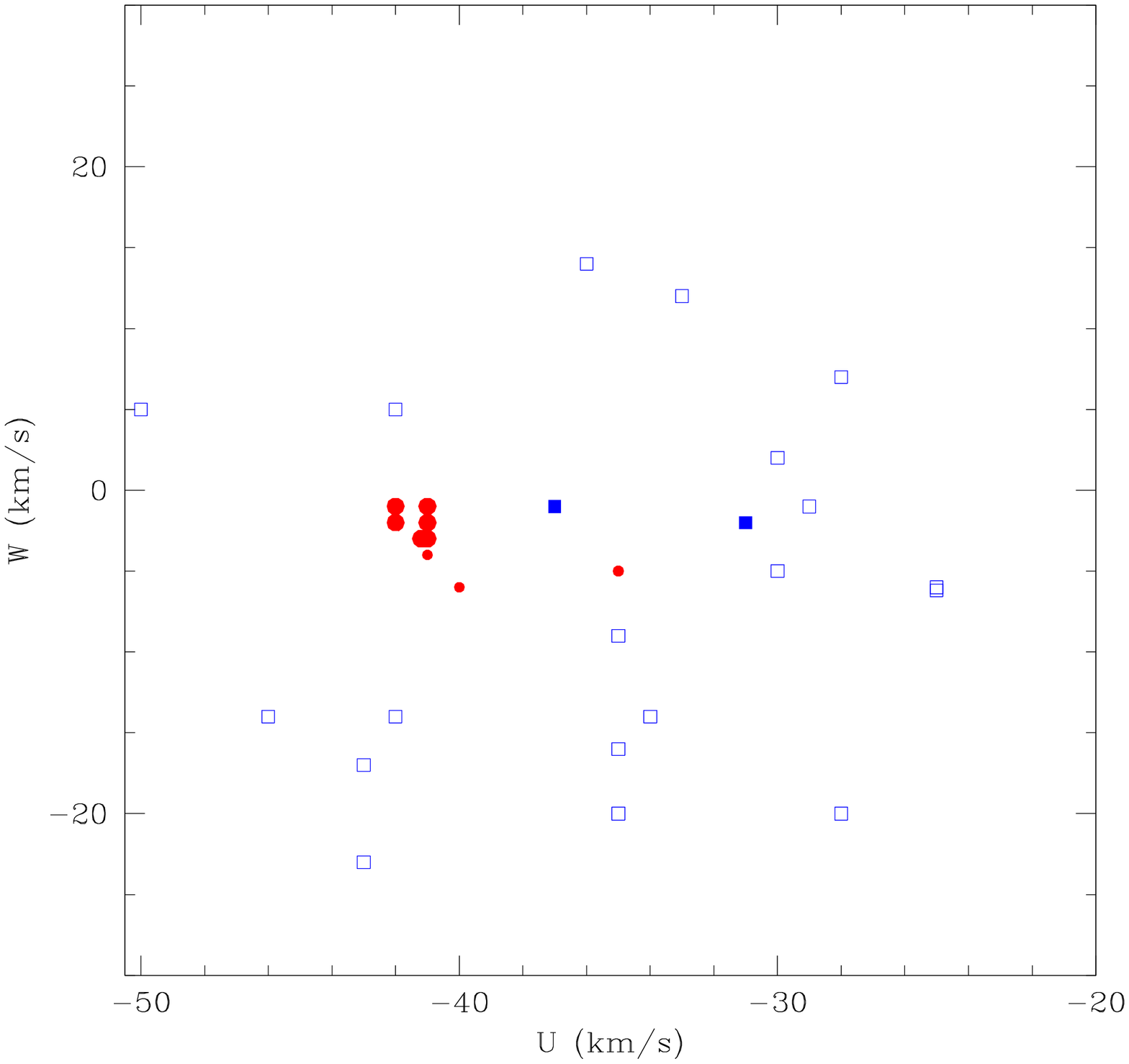}
\caption{Same as Fig.~\ref{Fig:UV} for the  $U-W$ plane.
}
\label{Fig:UW}
\end{figure}

\begin{figure}
\includegraphics[width=8cm]{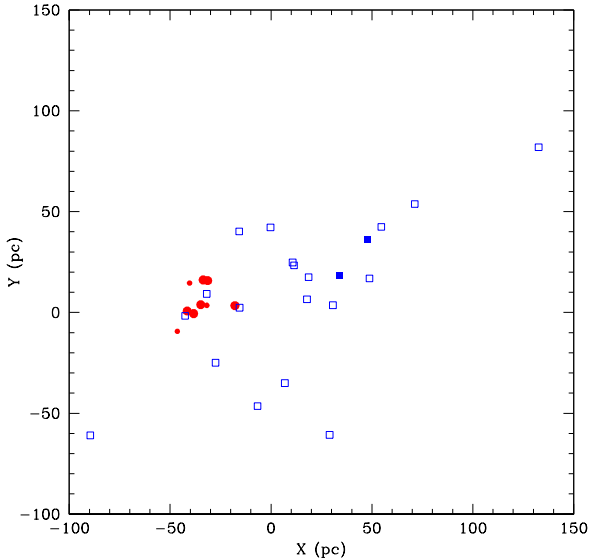}
\caption{
The spatial distribution of Hyades cluster and stream/field stars, 
projected on the $X-Y$ plane (with $X$ pointing
towards the galactic centre 
and $Y$ in the direction of the galactic rotation; distances on both
axes are expressed in parsecs). Symbols are as in Fig.~2.
}
\label{Fig:xy}
\end{figure}

\begin{figure}
\includegraphics[width=8cm]{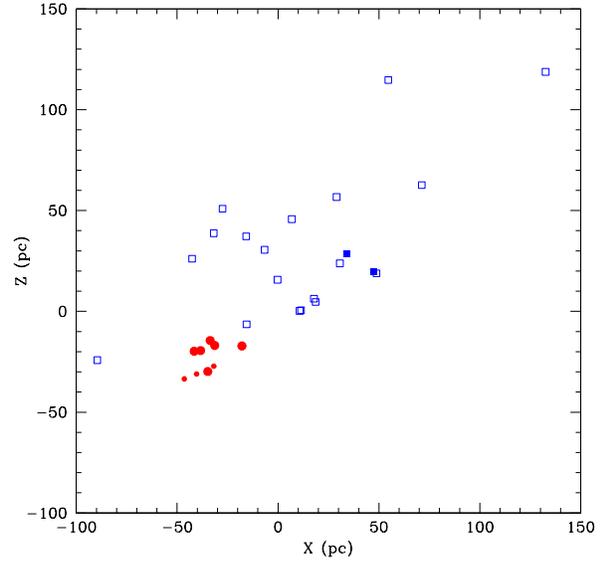}
\caption{Same as Fig.~\ref{Fig:xy} for the $X-Z$ plane.}
\label{Fig:xz}
\end{figure}

The full stellar sample is listed in Table~\ref{Tab:sample}, the distance and spatial velocities  are taken  from Holmberg et al. (2009), the masses from Holmberg et al. (2007), and the rotation velocity from Nordstr\"{o}m et al. (2004). Positions in the Galaxy and velocities are displayed in
Figs.~\ref{Fig:UV} to \ref{Fig:xz}. As each
star has been observed individually, the sample is relatively small,
comprising 21 stars randomly chosen in the Hyades box, 5 sure
members of the Hyades cluster, and 4 possible members. This small sample size nevertheless already allows us to draw a number of secure conclusions, as we shall see in the following sections, and paves the way towards more detailed analyses of much larger samples with future multi-fibre spectrographs such as the High Resolution Multi-Object Spectrograph of the Anglo-Australian Telescope, which like  the spectrometer used by us also is called HERMES (Barden et al. 2008).


%\newpage

\renewcommand{\baselinestretch}{1.0}
\begin{table} 
\caption[]{\label{linelist}
Line list. An asterisk (*) after the wavelength means that hyperfine structure and isotopic shifts have been taken into account for 
the considered line. References to the adopted $\log gf$ are given in the last column and the corresponding  reference is listed at the bottom of the table.}
\begin{tabular}{lccccclccr}
\hline
Wavelength & Element & $\chi$ & $\log gf$ & Ref\\
(\AA)  &        & (eV)\\
\hline
6707.7561 &Li I & 0.000 &-0.428 & (7)\\
6707.7682 &Li I &0.000 &-0.206 & (7)\\
6707.9066 &Li I &0.000 &-1.509 & (7)\\
6707.9080 &Li I &0.000 &-0.807 & (7)\\
6707.9187 &Li I &0.000 &-0.807 & (7)\\
6707.9200 &Li I &0.000 &-0.807& (7)\\
4497.7 &  Na I   &  2.104 &   -1.560  &(1)\\   
4668.6 &  Na I   &  2.104 &   -1.300  &(1) \\   
4751.8 &  Na I   &  2.104 &   -2.090  &(1) \\   
4982.8 &  Na I   &  2.104 &   -0.950  &(1) \\   
5148.8 &  Na I   &  2.102 &   -2.060  &(1) \\   
5682.6 &  Na I   &  2.102 &   -0.700  &(1) \\   
5688.2 &  Na I   &  2.104 &   -0.450  &(1) \\   
5889.9 &  Na I   &  0.000 &    0.117  &(1) \\   
5895.9 &  Na I   &  0.000 &   -0.184  &(1) \\   
6154.2 &  Na I   &  2.102 &   -1.560  &(1) \\   
6160.8 &  Na I   &  2.104 &   -1.260  &(1) \\   
5711.1 &  Mg II  &  4.346 &   -1.810  &(1) \\   
6318.7 &  Mg II  &  5.108 &   -1.970  &(1) \\   
6319.2 &  Mg II  &  5.108 &   -2.165  &(1) \\   
7387.7 &  Mg II  &  5.753 &   -1.100  &(1) \\   
7691.6 &  Mg II  &  5.753 &   -0.783  &(1) \\   
5079.7 &  Fe I   &  0.990 &   -3.220  &(1) \\   
5150.8 &  Fe I   &  0.990 &   -3.003  &(1) \\   
5151.9 &  Fe I   &  1.010 &   -3.322  &(1) \\   
5322.0 &  Fe I   &  2.280 &   -2.840  &(1) \\   
5400.5 &  Fe I   &  4.370 &   -0.160  &(1) \\   
5811.9 &  Fe I   &  4.140 &   -2.430  &(1) \\   
5853.1 &  Fe I   &  1.490 &   -5.280  &(1) \\   
5855.1 &  Fe I   &  4.610 &   -1.478  &(1) \\   
5856.1 &  Fe I   &  4.290 &   -1.328  &(1) \\   
5858.8 &  Fe I   &  4.220 &   -2.260  &(1) \\   
5927.8 &  Fe I   &  4.650 &   -1.090  &(1) \\   
5933.8 &  Fe I   &  4.640 &   -2.230  &(1) \\   
5956.7 &  Fe I   &  0.860 &   -4.605  &(1) \\   
5969.6 &  Fe I   &  4.280 &   -2.730  &(1) \\   
6019.4 &  Fe I   &  3.570 &   -3.360  &(1) \\   
6027.1 &  Fe I   &  4.080 &   -1.089  &(1)   \\
6054.1 &  Fe I   &  4.370 &   -2.310  &(1)   \\
6105.1 &  Fe I   &  4.550 &   -2.050  &(1)   \\
6151.6 &  Fe I   &  2.180 &   -3.299  &(1)   \\
6157.7 &  Fe I   &  4.080 &   -1.110  &(1)   \\
6159.4 &  Fe I   &  4.610 &   -1.970  &(1)   \\
6165.4 &  Fe I   &  4.140 &   -1.474  &(1)  \\
6173.3 &  Fe I   &  2.220 &   -2.880  &(1)   \\
6311.5 &  Fe I   &  2.830 &   -3.141  &(1)   \\
6315.8 &  Fe I   &  4.080 &   -1.710  &(1)   \\
6322.7 &  Fe I   &  2.590 &   -2.426  &(1)   \\
6355.0 &  Fe I   &  2.850 &   -2.350  &(1)   \\
6380.7 &  Fe I   &  4.190 &   -1.376  &(1)   \\
6392.5 &  Fe I   &  2.280 &   -4.030  &(1)   \\
6481.9 &  Fe I   &  2.280 &   -2.984  &(1)   \\
6498.9 &  Fe I   &  0.960 &   -4.699  &(1)  \\
6518.4 &  Fe I   &  2.830 &   -2.460  &(1)   \\
6574.2 &  Fe I   &  0.990 &   -5.023  &(1)   \\
6593.9 &  Fe I   &  2.430 &   -2.422  &(1)   \\
6597.6 &  Fe I   &  4.800 &   -1.070  &(1)   \\
6609.1 &  Fe I   &  2.560 &   -2.692  &(1)   \\
\hline
\end{tabular}
\end{table}
\addtocounter{table}{-1}
\begin{table} 
\caption[]{Continued.}
\begin{tabular}{lccccclccr}
\hline
Wavelength & Element & $\chi$ & $\log gf$ & Ref\\
(\AA)  &        & (eV)\\
\hline
6627.5 &  Fe I   &  4.550 &   -1.680  &(1)   \\
6710.3 &  Fe I   &  1.490 &   -4.880  &(1)   \\
6716.2 &  Fe I   &  4.580 &   -1.920  &(1)   \\
6725.4 &  Fe I   &  4.100 &   -2.300  &(1)   \\
6726.7 &  Fe I   &  4.610 &   -0.829  &(1)  \\
6750.2 &  Fe I   &  2.420 &   -2.621  &(1)   \\
6752.7 &  Fe I   &  4.640 &   -1.204  &(1)   \\
6793.3 &  Fe I   &  4.080 &   -2.326  &(1)   \\
6806.8 &  Fe I   &  2.730 &   -3.210  &(1)   \\
6810.3 &  Fe I   &  4.610 &   -0.986  &(1)   \\
6841.3 &  Fe I   &  4.610 &   -0.750  &(1)   \\
6843.7 &  Fe I   &  4.550 &   -0.930  &(1)   \\
4491.4 &  Fe II  &  2.860 &   -2.700  &(1)   \\
4508.3 &  Fe II  &  2.860 &   -2.210  &(1)   \\
4620.5 &  Fe II  &  2.830 &   -3.240  &(1)   \\
5197.6 &  Fe II  &  3.230 &   -2.230  &(1)   \\
5264.8 &  Fe II  &  3.230 &   -3.120  &(1) \\   
5325.6 &  Fe II  &  3.220 &   -3.220  &(1) \\   
5414.1 &  Fe II  &  3.220 &   -3.620  &(1) \\   
5425.3 &  Fe II  &  3.200 &   -3.210  &(1) \\   
6149.3 &  Fe II  &  3.890 &   -2.720  &(1) \\   
4687.8 &  Zr I   &  0.730 &    0.550  &(1) \\   
4739.5 &  Zr I   &  0.651 &    0.230  &(1) \\   
4815.1 &  Zr I   &  0.651 &   -0.530  &(1) \\   
4815.6 &  Zr I   &  0.604 &   -0.030  &(1) \\ 
6134.6 &  Zr I   &  0.000 &   -1.280  &(1) \\   
3714.8 &  Zr II  &  0.527 &   -0.930  &(1) \\   
3836.8 &  Zr II  &  0.559 &   -0.060  &(1) \\   
4151.0 &  Zr II  &  0.802 &   -0.992  &(1) \\   
4209.0 &  Zr II  &  0.713 &   -0.460  &(1) \\   
4211.9 &  Zr II  &  0.527 &   -1.083  &(1) \\   
4317.3 &  Zr II  &  0.713 &   -1.380  &(1) \\   
4379.7 &  Zr II  &  1.532 &   -0.356  &(1) \\   
4443.0 &  Zr II  &  1.486 &   -0.330  &(1) \\   
4497.0 &  Zr II  &  0.713 &   -0.860  &(1) \\   
4629.1 &  Zr II  &  2.490 &   -0.590  &(1) \\     
5112.3 &  Zr II  &  1.665 &   -0.590  &(1) \\   
5350.1 &  Zr II  &  1.827 &   -1.240  &(1) \\  
6498.8 &  Ba I   &  1.190 &    0.580  &(1) \\    
3891.8 &  Ba II  &  2.512 &    0.295  &(2) \\   
4130.7* &  Ba II  &  2.722 &    0.525  &(2) \\   
4166.0 &  Ba II  &  2.722 &   -0.433  &(2) \\   
4554.0* &  Ba II  &  0.000 &    0.140  &(2) \\   
4934.1* &  Ba II  &  0.000 &   -0.157  &(2) \\   
5853.7* &  Ba II  &  0.604 &   -0.909  &(2) \\   
6141.7* &  Ba II  &  0.704 &   -0.030  &(2) \\   
6496.9* &  Ba II  &  0.604 &   -0.406  &(2) \\   
3988.5* &  La II  &  0.403 &   -0.280  &(3) \\   
3995.7* &  La II  &  0.173 &   -0.060  &(3) \\   
4042.9 &  La II  &  0.927 &    0.290  &(3) \\   
4086.7* &  La II  &  0.000 &   -0.070  &(3) \\   
4123.2* &  La II  &  0.321 &    0.130  &(3) \\   
4322.5* &  La II  &  0.173 &   -0.930  & (3)\\   
4333.8 &  La II  &  0.173 &   -0.060  & (3)  \\ 
4522.4 &  La II  &  0.000 &   -1.200  &(3)   \\ 
4525.3 &  La II  &  1.946 &   -0.376  &(3)   \\ 
4619.9 &  La II  &  1.754 &   -0.122  &(3)   \\ 
4662.5* &  La II  &  0.000 &   -1.240  &(3)   \\ 
4716.4 &  La II  &  0.772 &   -1.210  &(3)   \\ 
4804.0 &  La II  &  0.235 &   -1.490  &(3)   \\ 
5805.8 &  La II  &  0.126 &   -1.560  &(3)   \\ 
6390.5* &  La II  &  0.321 &   -1.410  &(3)   \\ 
4115.4 &  Ce II  &  0.924 &    0.100  &(4)   \\ 
\hline
\end{tabular}
\end{table}
\addtocounter{table}{-1}
\begin{table} 
\caption[]{Continued.}
\begin{tabular}{lccccclccr}
\hline
Wavelength & Element & $\chi$ & $\log gf$ & Ref\\
(\AA)  &        & (eV)\\
\hline
4137.7 &  Ce II  &  0.516 &    0.440  &(4)   \\ 
4562.4 &  Ce II  &  0.478 &    0.230  &(4)   \\ 
4628.2 &  Ce II  &  0.516 &    0.200  &(4)   \\ 
5274.2 &  Ce II  &  1.044 &    0.150  &(4)   \\ 
4109.5 &  Nd II  &  0.321 &    0.350  &(5)   \\ 
4446.4 &  Nd II  &  0.205 &   -0.350  &(5)   \\ 
5319.8 &  Nd II  &  0.550 &   -0.140  &(5)   \\ 
3724.9* &  Eu II  &  0.000 &   -0.090  &(6)   \\ 
4129.7* &  Eu II  &  0.000 &    0.220  &(6)   \\ 
4205.0* &  Eu II  &  0.000 &    0.210  &(6)   \\ 
6049.5* &  Eu II  &  1.279 &    0.800  &(6)   \\ 
6645.1* &  Eu II  &  1.380 &    0.120  &(6)   \\ 
\hline
\end{tabular} 

(1) VALD Database (2) Masseron (2006) (3) Lawler et al. (2001a) (4) Zhang et al. (2001) (DREAM database) \\
(5) Den Hartog et al. (2003)  (6) Lawler et al. (2001b)  (7) Hobbs et al. (1999)\\
\end{table}
\renewcommand{\baselinestretch}{2.0}

\section{Observations}

The selected stars were observed in May 2009 with the Mercator 1.2m
telescope (Katholieke Universiteit Leuven, Roque de los Muchachos
Observatory, La Palma, Spain), using the fiber-fed, high-resolution
spectrograph HERMES/Mercator (Raskin et al.  2011).  
The spectra cover the wavelength range 380 to 900~nm, with a mean resolving power 
$ \lambda / \Delta \lambda $ = 84600, and a mean S/N of about
125 (the S/N ratio for each target star is listed in
Table~\ref{Tab_param}). The data were reduced using the automatic pipeline for the HERMES/Mercator
spectrograph, complemented by IRAF
packages for continuum definition and Doppler correction. 

\section{Determination of the stellar parameters}

\begin{figure}
\includegraphics[width=6cm,angle=270]{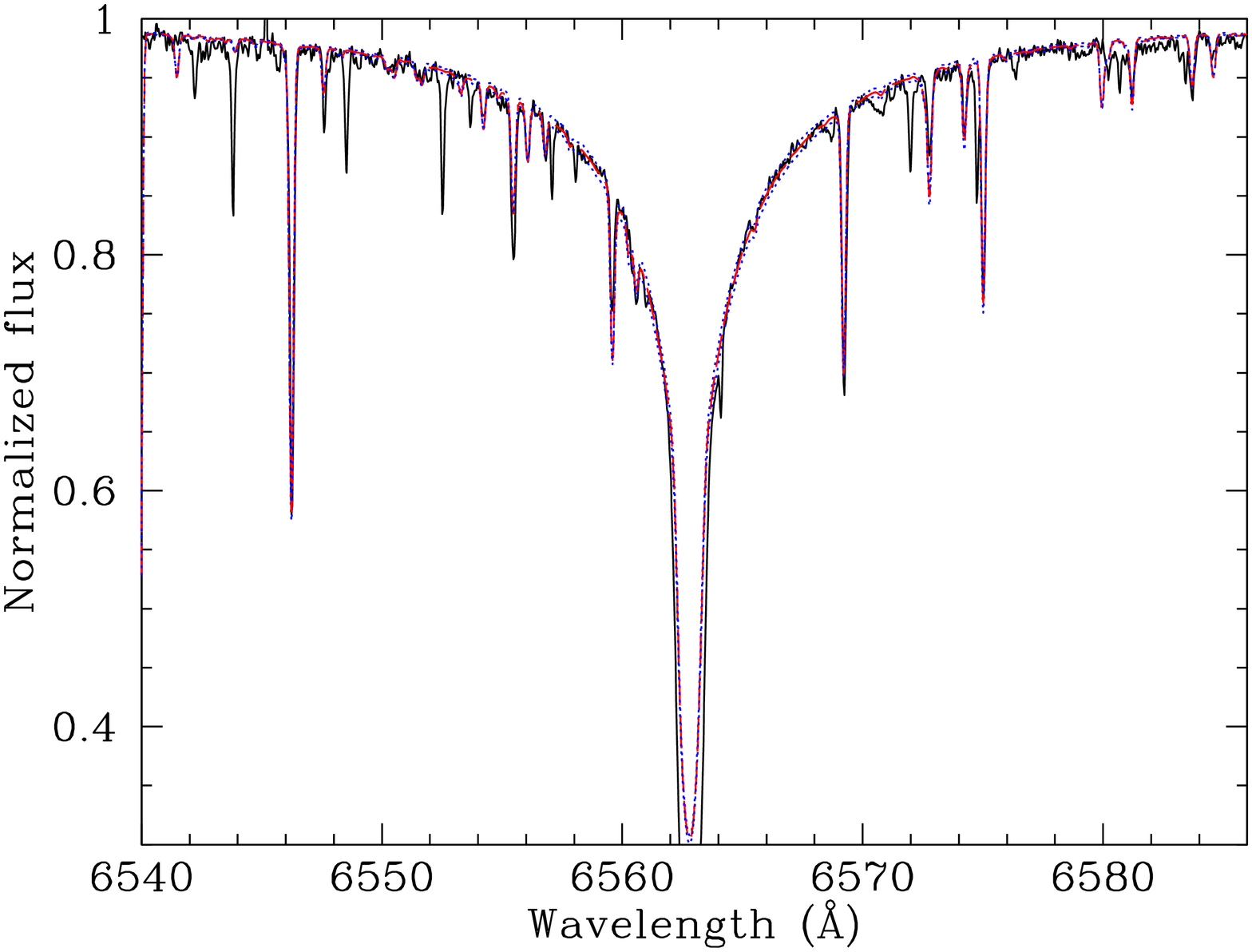}
\caption{\label{Fig:Ha}
 Fit of the H$\alpha$ line profile of HD~103891, for the adopted $T_{\rm eff}$ of 5900~K (red solid line) $\pm50$~K (blue dotted lines).
}
\end{figure}

We have adopted the new MARCS model atmospheres (Gustafsson et al. 2008)
and interpolated within the grid whenever necessary.
The stellar parameters were obtained in the  following manner. The effective
temperature was derived by
requiring that the abundances derived from iron lines (as listed
in Table~\ref{linelist}) exhibit no trend
with excitation potential. Gravity was obtained through ionisation
balance, and
microturbulence was set by requiring the absence of a trend between
abundances
and reduced equivalent widths. It was moreover checked that the wings of
the H$\alpha$ line are correctly reproduced (Fig.~\ref{Fig:Ha}).

\begin{figure}
\includegraphics[width=6cm,angle=270]{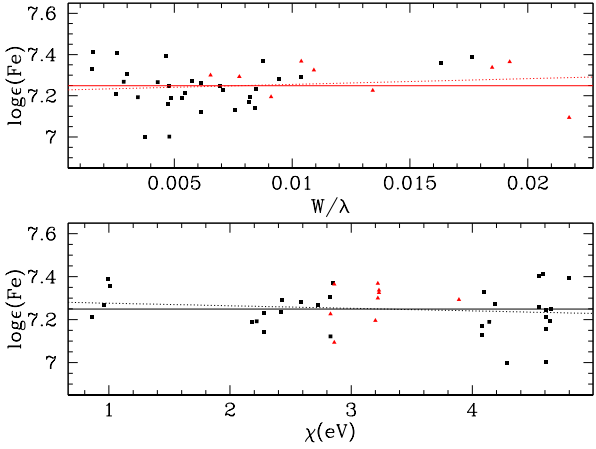}
\caption{\label{Fig:EW_EP}
[Fe~I / H] (black filled squares) and [Fe~II / H] (red filled triangles) against the line equivalent widths and excitation potentials, for star HD~103891. The solid line depicts the 
adopted abundance, whereas the dotted line is the least-square fit 
through the data points, with slopes of  $2.80\pm3.27$ and $-0.012\pm0.014$ for the upper and lower panels respectively, thus consistent with a zero-slope within the error bars.
}
\end{figure}

To illustrate the procedure, Fig.~\ref{Fig:EW_EP} displays [Fe~I / H] and [Fe~II / H]
against the line 
equivalent widths and excitation potentials, for star HD~103891. The difference  $T_{\rm eff}$(H$\alpha$)$-  T_{\rm eff}$(Fe~I) averages to $-20$~K, with a standard deviation of 33~K for our complete sample.

\begin{figure}
\includegraphics[width=9cm]{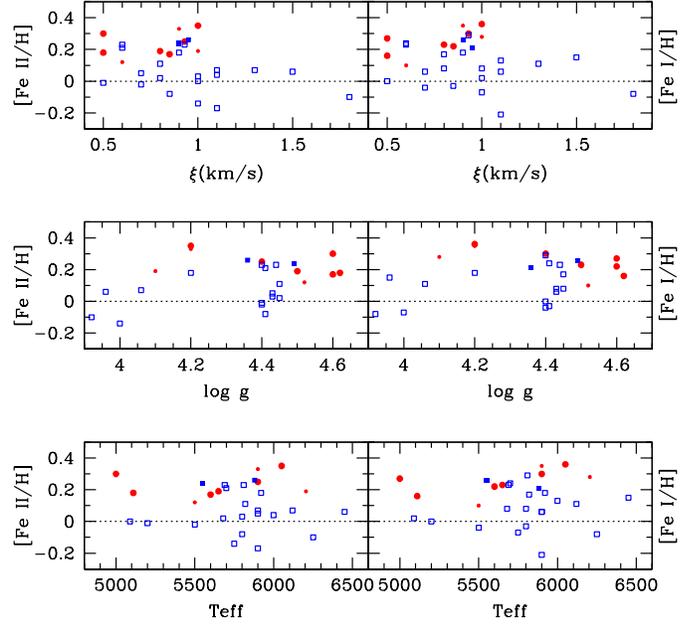}
\caption{Metallicities as a function of the adopted stellar parameters
  $T_{\rm eff}, \log g$ and microturbulence $\xi$ (in km/s). 
Symbols are as in Fig.~2. 
}
\label{Fig:FeH}
\end{figure}

The adopted atmospheric parameters are
listed in Table~\ref{Tab_param}, where the stars are already  grouped
according to their abundances and kinematics (the separation procedure
will be discussed in Sect.~\ref{Sect:evap}). The metallicity values
listed
in Table~\ref{Tab_param} and plotted in Fig.~\ref{Fig:FeH} are
normalized by the (solar) value $\log \epsilon({\rm Fe}) = 7.45$ (Asplund et al. 2005). 
In this table, we also list the S/N ratio of the spectra, and the
line-to-line abundance scatter $\sigma$. 
Fig.~\ref{Fig:FeH} checks for the absence of any trend between  [Fe~I,
Fe~II~/~H] and the atmospheric parameters 
microturbulence $\xi$, gravity $\log g$ and effective temperature
$T_{\rm eff}$. 

 The solar metallicity has been derived from the same set of lines as for the program stars, with  the following solar-model parameters: $T_{\rm eff} =  5777$~K, $\log g =  4.44$, [Fe/H] =  0.00, $\xi = 1.0~\kms$, 
and using the solar spectrum from Neckel (1999). The solar metallicity [Fe~I/H]$_\odot = 0.13\pm0.12$  listed in Table~\ref{Tab_param} is just $1\sigma$ off the expected null value.

For comparison, the atmospheric  parameters derived by Paulson et al. (2003) for the stars in common are listed as well in Table~\ref{Tab_param}, along
with $T_{\rm eff}$ from the fit of the H$\alpha$ wings. 
For the 9 stars in common with Paulson et al. (2003), our final stellar
parameters agree well with theirs, since  $T_{\rm eff}$(Paulson)$-  T_{\rm eff}$(Fe~I, this work) averages to 70~K with a standard deviation of 75~K (and a maximum difference  of 200~K  for HD~20430); $\log g$(Paulson)$-  \log g$(this work) averages to 0.05 with a standard deviation of 0.11 (and a maximum difference of 0.2~dex for HD~20430, HD~20439 and HD~26257). Although 
our metallicities may appear systematically larger than Paulson et al. ones in Table~\ref{Tab_param}, it must be noted that the normalisation value is different, since Paulson et al. (2003) have used $\log \epsilon_\odot$(H)$ = 7.64$ (as derived from their statement that their value of the solar metallicity differs by 0.14~dex from Grevesse \& Sauval one: $\log \epsilon_\odot$(H)$ = 7.50$). After allowing for this difference between the normalisation values, it turns out that our metallicities are on average 0.08~dex smaller than Paulson ones,  the largest difference being 0.14~dex for HD~20439, and the second largest 0.13~dex for HD~20430 and HIP~13806.

\begin{figure*}
\includegraphics[width=15cm]{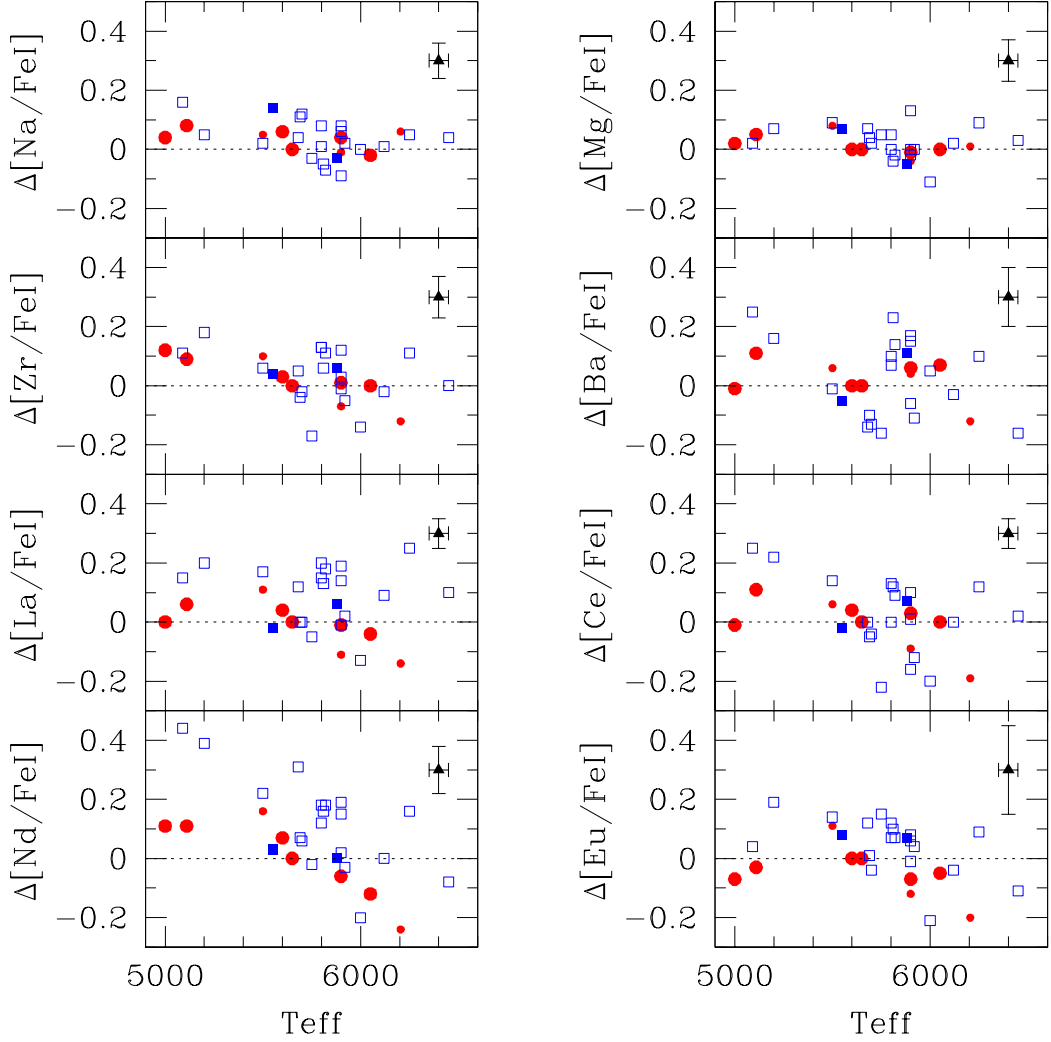}
\caption{Differential [X/Fe] ratios (relative to HD 26756) for Na, Mg, Zr, Ba, La, Ce, Nd and Eu against $T_{\rm eff}$. Symbols are as in
Fig.~2.  The error bars drawn in each panel represent 
the abundance uncertainties due to  uncertainties on the stellar
parameters (see Table~\ref{TabErr}). The dashed line corresponds to [X/Fe] = 0.
}
\label{diff}
\end{figure*}

\renewcommand{\baselinestretch}{1.0}

\begin{table*}
\caption[]{\label{Tab_param}
Stellar parameters. The [Fe/H] values from this work are normalized by  $\log \epsilon({\rm Fe}) = 7.45$ (Asplund et al. 2005), while those from Paulson et al. (2003) are normalized by $\log \epsilon({\rm Fe}) = 7.64$. 
}
\begin{tabular}{rcclcclcccrcccccc}
\hline
\hline
HD &  $T_{\rm eff}$ & $\log g$ & \multicolumn{2}{c}{[Fe I /H] $\sigma$}&
nlines& \multicolumn{2}{c}{[Fe II/H]  \hspace{\fill} $\sigma$} & nlines & $\xi$ &
S/N & $T_{\rm eff}$ & $T_{\rm eff}$ & $\log g$ & [Fe/H] & Membership$^a$\\
\cline{13-15}    & (K) & & & & & & &  & (km s$^{-1}$) & & H$\alpha$& \multicolumn{3}{c}{(Paulson et al. 2003)}\\
\hline
&& \multicolumn{7}{c}{\bf Hyades cluster} \\
\hline\\
&& \multicolumn{7}{c}{Sure members}& 
\medskip\\
18632   & 5000  &  4.60 &   0.27 &  0.16 & 39 &\phantom{-}0.30 &  0.18 & 7 &  0.5 & 107 & 5000 & 5000 & 4.6 & 0.18 & 1\\
19902   & 5600  &  4.60 &   0.22 &  0.09 & 35 &\phantom{-}0.17 &  0.12 & 8 &  0.85 & 107 & 5550 & 5600 & 4.5 & 0.09 & 1\\
26756   & 5650  &  4.50 &   0.23 &  0.11 & 42 &\phantom{-}0.19 &  0.09 & 9 &  0.8 & 118 & 5625 & 5650 & 4.5 & 0.06 & 1\\
26767   & 5900  &  4.40 &   0.30 &  0.10 & 38 &\phantom{-}0.25 &  0.09 & 9 &  0.9 & 91  & 5800 & 5900 & 4.4 & 0.12 & 1\\
HIP 13806& 5110 &  4.62 &   0.16 &  0.13 & 45 &\phantom{-}0.18 &  0.17 & 9 &  1.0 & 143& 5100 & 5200 & 4.6 & 0.10 & 1 \\
\medskip\\
&& \multicolumn{7}{c}{Possible members}& 
\medskip\\
20430   & 6050  &  4.20 &   0.36 &  0.08 & 32 &\phantom{-}0.35 &  0.08 & 9 &  1.0 & 128  & 6050 & 6250 & 4.4 & 0.30 & 5 \\
20439   & 5900  &  4.20 &   0.35 &  0.10 & 35 &\phantom{-}0.33 &  0.09 & 8 &  0.9 & 151  & 5900 & 6050 & 4.4 & 0.30 & 5 \\
26257   & 6205  &  4.10 &   0.28 &  0.10 & 31 &\phantom{-}0.19 &  0.11 & 8 &  1.0 & 156 & 6150 & 6300 & 4.3 & 0.11 & 4\\
HIP 13600& 5510 &4.52 &  0.10 &  0.07 & 35 &\phantom{-}0.12 &  0.12 & 9 &  0.6 & 147& 5500 & 5600 & 4.5 & 0.02 & 4 \\
\hline\\
&& \multicolumn{7}{c}{\bf Hyades stream} \\
\hline\\
%\hline
&& \multicolumn{7}{c}{Evaporated candidates}& 
\medskip\\
149028  & 5550  &  4.49 &   \phantom{-}0.26 &  0.10 & 41 & \phantom{-}0.24  &  0.12 & 9 &   0.9 & 106  & 5525   \\
162808  & 5880  &  4.36 &   \phantom{-}0.21 &  0.10 & 36 &  \phantom{-}0.26 &  0.08 & 9 &   0.95 & 91 & 5875  \\
&& \multicolumn{7}{c}{Stream/field}& 
\medskip\\
25680   & 5800  &  4.43 &   \phantom{-}0.08 &   0.11 & 40 & \phantom{-}0.03  & 0.10  & 9 & 1.0 & 175 & 5800  \\
42132   & 5090  &  3.00 &   \phantom{-}0.02 &   0.12 & 38 & \phantom{-}0.00  &  0.12 & 9 &  1.0& 175 & 5050  \\
67827   & 5900  &  3.90 &   \phantom{-}0.06 &  0.10 & 38 & \phantom{-}0.07 &  0.09 & 9 &   1.1  & 155 & 5875 \\
86165   & 5750  &  4.00 &              -0.07 &  0.16 & 35 &            -0.14  &  0.09 & 9 &   1.0 & 122 & 5825 \\
89793   & 5690  &  4.44 &   \phantom{-}0.23  &  0.10 & 44 & \phantom{-}0.23   &  0.12 & 9 &   0.6 & 84  & 5650 \\
90936   & 5915  &  4.20 &   \phantom{-}0.18 &  0.09 & 37 & \phantom{-}0.18  &  0.08 & 9 &   0.9 & 89 & 5900 \\
103891  & 5900  &  3.70 &              -0.21 &  0.10 & 31 &            -0.17  &  0.09 & 9 & 1.1 & 162 & 5900\\
108351  & 6120  &  4.06 &   \phantom{-}0.11 &   0.10 & 26 & \phantom{-}0.07   & 0.08 & 8 & 1.3 & 174 & 6075 \\ 
133430  & 5800  &  4.41 &              -0.03 &  0.09 & 35 &            -0.08  &  0.09 & 9 &   0.85 & 151 & 5775 \\
134694  & 6450  &  3.96 &   \phantom{-}0.15 &  0.09 & 21 &  \phantom{-}0.06 &  0.10 & 9 &   1.5 & 122  & 6375\\
142072  & 5810  &  4.40 &   \phantom{-}0.29 &   0.11 & 42 &  \phantom{-}0.23  &  0.10 & 8 &   0.9 & 115 & 5750\\
149285  & 6250  &  3.92 &              -0.08 &  0.09 & 13 &             -0.10 &  0.14 & 8 &   1.8 & 107 & 6250\\
151766  & 6000  &  3.50 &    \phantom{-}0.13 &   0.14 & 15 &  \phantom{-}0.04 &  0.18 & 8 &   1.1 & 121 & 6000\\
155968  & 5700  &  4.41 &   \phantom{-}0.24 &  0.10 & 43 & \phantom{-}0.21  &  0.10 & 9 &   0.6 & 93 & 5650\\ 
157347  & 5680  &  4.45 &   \phantom{-}0.08 &   0.10 & 39 &  \phantom{-}0.02 &  0.11 & 9 &   0.8 & 129 & 5650\\
171067  & 5500  &  4.40 &                   -0.04 &   0.08 & 30 &             -0.02 &  0.08 & 9 &   0.7 & 75 & 5525\\
180712  & 5900  &  4.43 &   \phantom{-}0.06 &   0.07 & 31 &  \phantom{-}0.05 &  0.09 & 8 &   0.7 & 100 & 5900\\
187237  & 5820  &  4.45 &   \phantom{-}0.17 &   0.10 & 34 &  \phantom{-}0.11  &  0.10 & 8 &   0.8 & 163 & 5800\\
189087  & 5205  &  4.40 &   \phantom{-}0.00 &  0.08 & 32 &             -0.01 &  0.13 & 8 &   0.5 & 94  & 5200\\
\hline\\
&& \multicolumn{7}{c}{\bf Sun}\\
\hline\\
Sun & 5777 & 4.44 & \phantom{-}0.13 &   0.12 & 39 &  \phantom{-}0.02  &  0.12 & 9 &   1.0&  \\
\hline
\end{tabular}

$^a$ The membership flag is from de Bruijne et al. (2001) according to
the following codes: (1) member based on proper motion and radial
velocity; (4) member based on proper motion and radial velocity but
rejected by Hoogerwerf \& Aguilar (1999); (5) member based
on proper motion and radial velocity but rejected by de Bruijne (1999)
and Hoogerwerf \& Aguilar (1999).  
\end{table*}

\renewcommand{\baselinestretch}{2.0}

\section{Abundances}

The choice of the elements studied was driven by the finding from previous abundance studies of the Hyades cluster by Paulson et al. (2003) and De Silva et al. (2006)  that the star-to-star differential abundance scatter for the elements Mg, Zr, Ba, La, Ce and Nd is especially small (at most 0.055~dex). To these elements, we have added Li, Na and the $r$-process element Eu.  
Our line list is given in Table~\ref{linelist}. Clean lines were first
selected on the Sun spectrum. Among those,  
some were discarded if they were too weak  and/or too severely blended in
most of the program stars. 
Elemental abundances for Na, Mg, Zr, Ba, La, Ce, Nd and Eu have been
derived from spectral-line synthesis, using the "turbospectrum" package (Alvarez \& Plez 1998), a program devoted to spectrum
synthesis in cool stars. The spectral-line-synthesis approach is well suited for complex lines with
hyperfine structure splitting (HFS) and/or isotopic shift (IS). 
The HFS and IS for Ba~II has
been taken from Masseron (2006), for La~II from Lawler et al. (2001a) 
and for Eu~II from Lawler et al. (2001b).  Concerning the
isotopic composition, 
we use the solar mixture (Grevesse \& Sauval, 1998), except for Li, for which we neglected the contribution from $^6$Li.    

For comparison, solar abundances have been derived using the same line list as for the target stars, 
with the solar-model parameters listed in Table~\ref{Tab_param}
and using the solar spectrum from Neckel (1999). Solar abundances normalized to $\log \epsilon$(H)$=12$ are listed in Table~\ref{Tab:sun}.

\begin{table}
\caption{\label{Tab:sun}
Solar elemental abundances, in the scale  $\log \epsilon$~(H)$=12$. Column 1 lists the considered ions, Col. 2 the meteoritic abundances from  Grevesse \& Sauval (1998), Cols. 2 and 3 the solar photospheric
standard abundances from Grevesse \& Sauval (1998) and Asplund et al. (2005), respectively. 
Our solar abundances are given in Col. 5. 
}
\begin{tabular}{rcccc}
\hline\\
Ion & Meteorites & \multicolumn{2}{c}{Photosphere} & This study  \\
   &    & GS98 & As05\\ 
\hline\\
Na I & 6.32 & 6.33 & 6.17 &  6.26 \\
Mg II & 7.58 & 7.58 & 7.53 & 7.56 \\
Fe I & 7.50 & 7.50 & 7.45 & 7.58  \\
Fe II & 7.50 & 7.50  & 7.45 & 7.47 \\
Zr I/II & 2.61 & 2.60 & 2.59 & 2.52\\
Ba II & 2.22 & 2.13 & 2.17 & 2.08 \\
La II & 1.22 & 1.17 & 1.13 & 1.10 \\
Ce II & 1.63 & 1.58 & 1.58 & 1.61 \\
Nd II & 1.49 & 1.50  & 1.45 & 1.38 \\
Eu II & 0.55 & 0.51 & 0.52 & 0.48 \\
\hline\\
\end{tabular}
\end{table}

Errors are expected to come from continuum placement, line fitting, and
uncertainties in stellar parameters and in atomic data. 
In order to reduce systematic errors due to the latter, we have
performed a line-by-line differential analysis relative to HD~26756, a
member of the 
Hyades cluster. With this differential approach, errors due to
uncertainties in the atomic data ($\log gf$) are avoided. The
differential 
abundances are given in Table~\ref{TabAb} together with the
line-to-line scatter ($\sigma$) and the number of lines used for each
element ($n$). Fig.~\ref{diff} presents  the differential [X/Fe] abundances (relative to HD 26756) for Na, Mg, Zr, Ba, La, Ce, Nd and Eu against $T_{\rm eff}$. Except perhaps for Nd (and to a lesser extent, Zr), there is no significant abundance trend  with $T_{\rm eff}$.

Errors due to continuum placement and line blends can be estimated
together with random errors
from the line-to-line scatter $\sigma$. 
For most of the lines, the uncertainties due to the 
line-fitting procedure are less than $\pm$0.02~dex, but for a few lines
in some stars it can be as high 
as $\pm$0.05~dex. 
For all elements the reported abundances are from an LTE analysis
although departures from LTE are expected for Na, Mg, Ba and Eu. 

In Table \ref{TabErr}, we list the uncertainties associated with the
stellar parameters, for two typical stars 
(HD~149285 and HD~189087). The adopted uncertainties for the parameters
are: $\Delta \log g = 0.1$~dex, 
$\Delta T_{\rm eff} = 50$~K and $\Delta$~[Fe/H]$ = 0.05$~dex. For the
microturbulent velocity, we have adopted 
$\Delta \xi = 0.10 \kms$, which is the minimum value for noticeable
changes in the slope of  [Fe/H] (Fig.~\ref{Fig:FeH}). The total
uncertainty 
($\Delta_{\rm err}$) has been derived assuming that the 
 errors coming from the uncertainties on the various atmospheric parameters
are not correlated (thus taking the square root of the quadratic
sum of the different error terms), and is listed in the last column of
Table~\ref{TabErr} for each element.
It turns out that it is Eu which has the largest sensitivity to errors on the stellar-model parameters, since $\Delta_{\rm err}({\rm Eu}) = 0.15$~dex as compared to less than 0.10~dex for all the other elements.

\renewcommand{\baselinestretch}{1.0}
%\begin{landscape}
\begin{table*}
\caption[]{\label{TabAb}
Abundance ratios, line-to-line scatter and number of lines used in the analysis. For HD~26756, we give the  
abundances normalized by the Asplund et al. (2005) solar abundances.
}
\tabcolsep 3pt
\begin{tabular}{rrrrrrrrrrrrrrrrrrrrrrrrrrrrrr}
\hline
\hline
HD/HIP & $\log\epsilon$(Li) & \multicolumn{2}{c}{[Na/Fe]     $\sigma$}& $n$ & \multicolumn{2}{c}{[Mg/Fe] $\sigma$} & $n$ &  \multicolumn{2}{c}{[Zr/Fe]    $\sigma$}& $n$ &  \multicolumn{2}{c}{[Ba/Fe] $\sigma$} & $n$ &  \multicolumn{2}{c}{[La/Fe]  $\sigma$}& $n$  & \multicolumn{2}{c}{[Ce/Fe] $\sigma$} & $n$ &  \multicolumn{2}{c}{[Nd/Fe]  $\sigma$} & $n$ &  \multicolumn{2}{c}{[Eu/Fe]  $\sigma$}& $n$ \\ 
\hline\\
\noalign{\bf Sure Hyades cluster members}\\
\hline\\
\noalign{Reference star, with [X/Fe] abundances normalized to Asplund et al. (2005) solar values}\\
\hline
26756     & 2.04 & -0.05 & 0.08 & 11 &  -0.08 & 0.17 & 11 & -0.16 & 0.14 & 10 & 0.12 & 0.13 & 6 & -0.17 & 0.07 & 8 & -0.01 & 0.33 &  5 & -0.16 & 0.09 & 3  & -0.10  & 0.04 & 3  \\ 
\hline\\
\noalign{Relative abundances [X/Fe] with respect to HD 26756}\\
18632      & $<0.4$ & 0.04 & 0.09 & 10  & 0.02 & 0.05 & 5 & 0.12 & 0.10 & 6 & -0.01 & 0.10 & 4 &   0.00   & 0.03  &  4  & -0.01 &  0.17   & 3 &  0.11 &   0.01  &2 &  -0.07 &  0.07 & 2  \\ 
19902      &  1.62 & 0.06 &  0.06 & 8 & 0.00 &0.05  & 5 &  0.03 & 0.11 & 10 &  0.00 & 0.06 &  5 &  0.04 & 0.09 & 7 &  0.04 & 0.01 &3&  0.07 &  0.03 & 3  & 0.00 & 0.07 & 2  \\ 
26767      & 2.66 & 0.04 &  0.07 & 9 & -0.01 & 0.09 & 5 & 0.01 & 0.15 & 6 &  0.06 & 0.07 &  5 &  -0.01 & 0.06 & 7 & 0.03 & 0.03 &3&  -0.06 &  0.06 & 3 & -0.07 & 0.02 &2  \\ 
HIP13806   &  $<0.4$ & 0.08 & 0.06 & 8 & 0.05 & 0.06 &5 &  0.09 & 0.14 & 5 &  0.11 & 0.07 &  4 & 0.06 & 0.06 & 5 &  0.11 & 0.16 &3& 0.11 & 0.10 & 3 & -0.03 & -- &1  \\ 
\hline\\
\noalign{\bf Possible Hyades cluster members} \\
\hline\\
20430      & 2.95 & -0.02 & 0.07 & 10 & 0.00 & 0.08 & 5 & 0.00 & 0.12 & 5 & 0.07 & 0.11 &  5 & -0.04 & 0.06 & 6 & 0.00 & 0.01 & 3 &  -0.12 &  0.01 & 3 & -0.05 &  0.02 &3  \\ 
20439      & 2.73 & -0.01 &  0.04 & 9 & -0.04 & 0.08  & 5& -0.07  & 0.11 & 6 & 0.04 & 0.14 &  5 & -0.11 & 0.06 & 7 &  -0.09 & 0.02 & 3&  -0.06 &  0.05 & 3 & -0.12 &  0.02 &2  \\ 
26257      & 2.72 & 0.06 & 0.06 & 9  & 0.01 & 0.10 &4 &  -0.12 & 0.13 & 6 & -0.12 & 0.08 &  5 & -0.14 &  0.08   & 7 & -0.19 & 0.00 &3&  -0.24 &  0.11 & 3  & -0.20 & 0.01 &2  \\ 
HIP13600   &  1.61 & 0.05 & 0.08 & 9 & 0.08 & 0.07 &5 &  0.10 & 0.13 & 8 &  0.06 & 0.11 &  6 &  0.11 & 0.07 & 8 &  0.06 & 0.12 &5& 0.16 &  0.01 & 3  & 0.11 & 0.10  &3   \\  
\hline\\
\noalign{\bf Evaporated candidates} \\
\hline\\
149028    & 1.35 & 0.14 & 0.08 & 11 & 0.07 & 0.08 &5 & 0.04 & 0.09 & 10& -0.05 & 0.07 & 5 &  -0.02 & 0.07 & 7  & -0.02  & 0.02 & 4 & 0.03 & 0.05 & 3  & 0.08 & 0.06 & 3\\
162808    & 2.66 & -0.03 & 0.06 & 10 &  -0.05 & 0.09 &5 & 0.06 & 0.13 & 6& 0.11 & 0.07 & 4 & 0.06 & 0.08 & 7  & 0.07  & -- & 1 & 0.00 & 0.09 & 3 & 0.07 & 0.06 & 2\\ 
%\hline\\ 
%&&& \multicolumn{15}{c}{\bf Stream stars possibly evaporated from the cluster} \\
%\hline\\
\hline\\
\noalign{\bf Stream/field stars} \\
\hline\\
25680    &  2.42 & 0.01 & 0.07  & 11 & 0.00 & 0.10 &5 & 0.13 & 0.09 & 8 & 0.07  & 0.08 &  6 & 0.15  & 0.03 & 8  & 0.00 & 0.15 & 5 & 0.12 & 0.09 & 3 & 0.07 &0.04 &3 \\
42132    &  $<0.40$ & 0.16 & 0.06 & 11 &  0.02 & 0.07 &5 &0.11 & 0.14 & 10 & 0.25  & 0.14 &  5 & 0.15  & 0.07 & 5  & 0.25 & 0.12 & 4 & 0.44 & 0.03 & 2 & 0.04 &0.08 &3 \\
67827     &  2.22 & 0.08 & 0.10 & 11 & -0.02 & 0.01 &3 & -0.01 & 0.14 & 8& -0.06 & 0.07 & 5 & -0.01 & 0.06 & 8  & -0.16  & 0.08 & 4 &0.02 & 0.04 & 3 & -0.01 &0.07 & 2 \\ 
86165    & 1.68 & -0.03 & 0.05 & 10 & 0.05 & 0.08 &5 & -0.17 & 0.09 & 5 & -0.16  & 0.08 &  5 & -0.05  & 0.06 & 8  & -0.22 & 0.10 & 4 & -0.02 & 0.04 & 3 & 0.15 & 0.03 &2 \\ 
89793    &  $<1.3$ & 0.11 & 0.07 & 11 & 0.04 & 0.06 &5 & -0.04 & 0.12 & 10 & -0.10  & 0.09 &  5 & 0.00  & 0.10 & 8  & -0.05 & 0.07 & 5 &  0.07 &  0.07 & 3 & 0.01 & 0.04& 3\\ 
90936     & 2.13 & 0.02 &0.09 & 9 & 0.00 & 0.09 &5 & -0.05 & 0.14 & 9& -0.11 & 0.10 & 6 &  0.02 & 0.05 & 8  & -0.12  & 0.10 & 5 & -0.03 &0.02  & 3 & 0.04 &0.06 & 2\\ 
103891   &  0.8 & 0.06 & 0.09 & 10 &  0.13 & 0.08 &3 & 0.12 & 0.10 & 7 & 0.15  & 0.07 &  5 & 0.19  & 0.04 & 8  & 0.01 & 0.11 & 5 &  0.19 & 0.03 & 3 & 0.08 &0.04  & 2 \\ 
108351   &  2.44 & 0.01 &0.07 & 10&  0.02 & 0.05 &2 & -0.02 & 0.15 & 6 &-0.03  & 0.06 &  4 &0.09  & 0.12 & 8  & 0.00 & 0.03 & 3 &  0.00 & 0.07 & 3 &-0.04 & 0.01& 2 \\
133430   &  2.08 & 0.08 & 0.05 & 10 &  0.05 & 0.09 &4 & 0.13 & 0.09 & 8 & 0.10  & 0.09 &  6 & 0.20  & 0.07 & 8  & 0.13 & 0.04 & 4 & 0.18 & 0.02 & 3 & 0.12 & 0.05 &2 \\
134694    &  $<1.6$ & 0.04 & 0.08 & 9 &  0.03 & 0.08 &4 & 0.00 & 0.10 & 6 &  -0.16 & 0.07 &  6 & 0.10 & 0.06 & 6  & 0.02  & 0.04 & 3 &  -0.08 & 0.17 & 3  & -0.11 &0.03 & 2\\  
142072   &  2.65 & -0.05 & 0.05 & 8 &  -0.04 & 0.08 &5 & 0.06 & 0.09 & 6 & 0.23  & 0.10 &  5 & 0.13  & 0.06 & 7  & 0.12 & 0.01 & 3 &  0.16 & 0.06 & 3 & 0.10 & 0.04& 3\\
149285   &  2.83 & 0.05 & 0.07 & 11 & 0.09 & 0.09 &5 & 0.11 & 0.15 & 6 & 0.10  & 0.13 &  5 & 0.25  & 0.05 & 7  & 0.12 & 0.12 & 5 & 0.16 & 0.11 & 3 & 0.09 &0.13&2 \\ 
151766   &  1.3 & 0.00 & 0.07 & 10 &  -0.11 & 0.095 &4 & -0.14 & 0.13 & 5 &0.05  & 0.18 &  5 & -0.13  & 0.09 & 8  & -0.20 & 0.06 & 4 & -0.20 &  0.06 & 3 & -0.21 &0.04 &2 \\ 
155968    & 1.24 & 0.12 & 0.07 & 10 & 0.02 & 0.06 &5 & -0.02 & 0.10 & 7&  -0.13 & 0.19 & 5 & 0.00 & 0.08 & 7  & -0.04  & 0.03 & 3 & 0.06 & 0.05 & 3 & -0.04 & 0.06 & 2\\
157347   &  $<0.8$ & 0.04 & 0.06 & 11 &  0.07 & 0.01 &3 & 0.05 & 0.08 & 9 &  -0.14 & 0.12 &  6 & 0.12  & 0.05 & 8  & 0.00 & 0.10 & 5 &  0.31 &  0.08 & 3 & 0.12 &0.04& 2\\ 
171067   &  $<0.8$ & 0.02 & 0.09 & 10 & 0.09 & 0.06 &5 & 0.06 & 0.08 & 8 & -0.01  & 0.09 &  6 & 0.17  & 0.06 & 8  & 0.14 & 0.01 & 3 & 0.22 & 0.05 & 3& 0.14 &0.05 &3 \\ 
180712   &  2.41 & -0.09 & 0.05 & 10 &  0.00  & 0.17 &5 & 0.03 & 0.10 & 6 & 0.17  & 0.05 &  5 & 0.14  & 0.05 & 7  & 0.10 & 0.06 & 5 &  0.15 & 0.01 & 3& 0.06 &0.05&3 \\ 
187237   &  2.18 & -0.07 & 0.06 & 10 &  -0.02 & 0.03 &4 & 0.11 & 0.10 & 8 & 0.14  & 0.08 &  5 & 0.18  & 0.05 & 8  & 0.09 & 0.04 & 3 &  0.18 & 0.00 & 3 & 0.07 &0.01&2 \\ 
189087   &  $<0.6$ & 0.05 & 0.08 & 9 &  0.07 & 0.05 &4 & 0.18 & 0.15 & 9 & 0.16  & 0.08 &  6 & 0.20  & 0.07 & 6  & 0.22 & 0.09 & 3 & 0.39 & 0.02 & 3 & 0.19 &0.10 &3 \\ 
\hline\\
%\noalign{\bf Sun}\\
%%Absolute abundances
%% &  0.00 &  0.40 & 0.13 &2 & -0.17 & 0.17 & 3 & 0.04  & 0.10 &  3 & 0.11  & 0.09 & 2  &  -0.19 & 0.01 & 2 & -0.10 &  -0.09 & 0.01 %&2 \\ 
%% Relative abundances
%Sun           & &  0.00 &  0.06 & 11 & -0.03&0.06 & 5 & -0.07 & 0.12 & 9 & -0.38  & 0.12 &  6 & -0.04  & 0.08 & 8  &  -0.15 & 0.06 & 5 & -0.05 &  0.06 %& 3 & -0.08 & 0.03 & 3 \\ 
%\hline  
\end{tabular} 
\end{table*}
\renewcommand{\baselinestretch}{2.0}

%\end{landscape}
\begin{table*}
\caption[]{\label{TabErr}
Errors in the abundances due to stellar parameter
uncertainties $\Delta \log g = 0.1$~dex, 
$\Delta T_{\rm eff} = 50$~K, $\Delta$~[Fe/H]$ = 0.05$~dex, 
$\Delta \xi = 0.10$~$\kms$. $\Delta_{\rm err}$ is the total error.
}
\begin{tabular}{lrrrrrrrrrrrr}
\hline
\hline
Element & \multicolumn{5}{c}{HD 149285} & \multicolumn{5}{c}{HD
  189087}\\  
\cline{2-6}\cline{8-12}
  & \multicolumn{1}{c}{$\Delta T_{\rm eff}$} &  \multicolumn{1}{c}{$\Delta \log g$} &  \multicolumn{1}{c}{$\Delta$ [Fe/H]}&  \multicolumn{1}{c}{$\Delta\xi$} & $\Delta_{\rm err}$
&&  \multicolumn{1}{c}{$\Delta T_{\rm eff}$} &  \multicolumn{1}{c}{$\Delta \log g$} & \multicolumn{1}{c}{$\Delta$ [Fe/H]} &  \multicolumn{1}{c}{$\Delta\xi$} & $\Delta_{\rm err}$ \\
\hline
$\Delta$[Na/Fe]   &  0.05 &  0.03 &  0.03 &  0.02  & 0.06 &&  0.06  & 0.02 &  0.01 &  0.01  & 0.06 \\
$\Delta$[Mg/Fe]   &  0.00 & -0.02 & -0.03 & -0.02  & 0.04 &&  0.00  & -0.02& -0.04 & -0.06  & 0.07 \\
$\Delta$[Zr/Fe]   &  0.01 &  0.03 & -0.03 & -0.02  & 0.05 &&  -0.02 & 0.01 & -0.06 &  -0.02 & 0.07 \\
$\Delta$[Ba/Fe]   &  0.02 &  0.01 &  0.05 &  0.07  & 0.09 &&  0.06  & 0.04 &  0.05 &  0.04  & 0.10 \\
$\Delta$[La/Fe]   &  0.02 &  0.03 &  0.02 &  0.00  & 0.04 &&  0.02  & 0.02 &  0.05 &  0.00  & 0.05 \\
$\Delta$[Ce/Fe]   &  0.02 &  0.03 &  0.04 &  0.03  & 0.06 &&  0.03  & 0.03 &  0.03 &  0.01  & 0.05 \\
$\Delta$[Nd/Fe]   &  0.02 &  0.02 &  0.06 &  0.04  & 0.07 &&  0.03  & 0.04 &  0.06 &  0.02  & 0.08 \\
$\Delta$[Eu/Fe]   & -0.07 & -0.04 & -0.12 & -0.08  & 0.16 && -0.05  &-0.06 & -0.11 & -0.07  & 0.15 \\
\hline
\end{tabular} 
\end{table*}

\section{Tagging the populations}
\label{Sect:evap}
\begin{table*}
\caption[]{\label{deviations}
Results form the chemical tagging based on Fe (relative to HD~26756) and the heavy elements (see text for details). Stars are ordered by increasing $F2(2)$ (see Eq.~\ref{Eq:F2}). The
column labelled `Member' is based on radial-velocity and proper-motion criteria (see Table~\ref{Tab_param}
for details). The horizontal line separates, in the right column, the sure Hyades members from the possible members, and
in the left column, the evaporated members from the stream/field stars.
}
\begin{tabular}{lrrrlrrrc}
\hline
\hline
Stream &$F2(1)$ &$F2(2)$ & [Fe~I/H] & Cluster & $F2(1)$ &$F2(2)$ & [Fe~I/H] & Member\\
\hline
                 &         &        &          &HD 26767  &-1.22 &-1.24 & 0.08 & 1\\
                 &         &        &          & HD 19902  &-0.76 &-1.08 & 0.00 & 1\\
                  &        &        &           & HD 26756  &-0.57 &-0.88 & 0.00 & 1\\
HD 162808 & 0.90 & 0.65 & -0.02 &  HD 20430  &0.16 &0.54 & 0.14 & 5\\
HD 149028 & 0.93 & 0.59 & 0.02    & HD 18632  &0.50 &0.16 & 0.04 & 1\\
                 &        &        &            & HIP 13806 &1.82&1.99 & -0.09 & 1\\ 
\cline{1-9} 
HD 108351 & 2.00& 2.19 & -0.10   & HIP 13600 &1.94 &2.48 & -0.14 & 4\\ 
HD 89793  & 2.77 & 2.53 & 0.00  \\
HD 155968 & 3.22 & 2.97 & 0.01    \\
HD 25680  & 3.64 & 3.96 & -0.14   & HD 20439  &3.99 &4.00 & 0.14 & 5\\ 
HD 157347 & 3.96 & 4.27 & -0.14 \\  
HD 90936  & 4.11 & 3.97 & -0.04 \\
HD 180712 & 4.12 & 4.53 & -0.16\\
HD 67827  & 4.19 & 4.58 & -0.16 \\ 
HD 134694 & 4.23 & 4.14 & -0.06\\  
HD 171067 & 4.55 & 5.50 & -0.25\\
HD 187237 & 4.91 & 4.83 & -0.06 \\  
HD 142072 & 4.94 & 4.77 & 0.06 \\ 
HD 103891 & 4.99 & 7.27 & -0.42\\       
HD 133430 & 5.42 & 6.21 & -0.25\\
HD 149285 & 6.55 & 7.41 & -0.28 \\
HD 151766 & 6.69 & 6.65 & -0.06  \\
HD 42132  & 7.14 & 7.53 & -0.20\\ 
HD 189087 & 7.20 & 7.76 & -0.24 & HD 26257  &7.22 &7.13 & 0.06 & 4\\
HD 86165  & 7.60 & 8.35 & -0.28\\
\medskip\\
\hline
\end{tabular} 
\end{table*}
 
\renewcommand{\baselinestretch}{1.0}
\begin{table*}
\caption[]{\label{meanvalues_delta1}
For each group (cluster, evaporated candidates, possible cluster, and stream/field, with the 
membership criterion based on $F2$; see Sect.~6) and each
chemical element considered, the table lists the mean ($\pm$ error on the mean),
standard deviation $\sigma_*$, and root-mean-square $s_{\rm lines}$ of
the line-to-line scatter. The differential  abundance scatter $\sigma_*$(P03) and $\sigma_*$(DS) obtained by Paulson et al. (2003) and De Silva et al. (2006; their Table~5), respectively,  from a differential abundance analysis of an extensive sample of 46 Hyades stars are listed as well for comparison.  The true scatter of the sample
($\sigma_{*,{\rm ext}}$) is obtained from $(\sigma^2_* - s^2_{\rm lines})^{1/2}$.
All abundances are differential with respect to HD~26756.
 The last column lists the [X/Fe I] (or [Fe I, II/H]) abundances of the reference star HD~26756, relative to the Asplund et al. (2005) solar abundances. 
}
\begin{tabular}{lllllllll}
\hline
\hline
         & Cluster  & Evaporated   & Possible    &Stream/field & HD 26756 \% solar\\
        &           &    candidates     & cluster     & population &  (Asplund et al. 2005) \\
\hline
 $N$       & 6      & 2           & 3                  & 19 \\  
 $F2$& $< 2$ &$< 2$& $\ge2$ &  $\ge 2$\\  
\hline
\multicolumn{2}{l}{Abundance}\\
\%HD~26756\medskip\\
\hline
$\langle\Delta$[Na/Fe]$\rangle$ & 0.03$\pm$0.01 & 0.05$\pm$0.06 & 0.03$\pm$0.02  & $0.03\pm0.01$ & -0.05 \\ %& 0.00\\
$\sigma_*(\Delta$[Na/Fe])       & 0.032                  & 0.09   & 0.03 & 0.06 \\
$\sigma_*$(P03)                      & 0.06 \\
$s_{\rm lines}(\Delta$[Na/Fe])   & 0.06          & 0.07 & 0.06 &  0.07 \\
$\sigma_{*,{\rm ext}}$           & $<0.03$       & 0.05  &  $<0.03$ & $<0.06$\\
\hline\\
$\langle\Delta$[Mg/Fe]$\rangle$ & 0.01$\pm$0.01 & $0.01\pm0.04$ & 0.02$\pm$0.03 & 0.03$\pm$0.01 & -0.08 \\ %& 0.03\\
$\sigma_*(\Delta$[Mg/Fe])       & 0.021          & 0.06 & 0.05   & 0.05 \\
$\sigma_*$(P03)                      & 0.04 \\
$s_{\rm lines}(\Delta$[Mg/Fe])   & 0.06          & 0.09 & 0.09 &  0.08 \\
$\sigma_{*,{\rm ext}}$           & $<0.02$       & $<0.06$ & $<0.05$   & $<0.05$\\
\hline \\
$\langle\Delta$[Fe I/H]$\rangle$& $0.03\pm$0.03 & $0.00\pm0.01$ &0.02$\pm$0.07 & $-0.14\pm$ 0.03 & 0.23 \\ %& 0.10\\
$\sigma_*(\Delta$[Fe I/H])      & 0.072          & 0.02 & 0.12    & 0.12 \\
$\sigma_*$(P03)                      & 0.05 \\
$s_{\rm lines}(\Delta$[Fe I/H])  & 0.08          & 0.06 & 0.06   & 0.09 \\
$\sigma_{*,{\rm ext}}$           & $<0.07$       & $<0.02$ & 0.10   & 0.08 \\
\hline\\
$\langle\Delta$[Fe II/H]$\rangle$& 0.05$\pm$0.03& $0.06\pm0.01$ &0.03$\pm$0.05 & -0.15$\pm$0.03 & 0.19 \\ % & 0.17\\
$\sigma_*(\Delta$[Fe II/H])     & 0.068          & 0.01 & 0.09  & 0.11 \\
$s_{\rm lines}(\Delta$[Fe II/H]) & 0.07          & 0.08 & 0.06  & 0.08 \\
$\sigma_{*,{\rm ext}}$           &0.00       & $<0.01$ & 0.07    & 0.07\\ 
\hline \\
$\langle\Delta$[Zr/Fe]$\rangle$ & 0.04$\pm$0.02 & 0.05$\pm$0.01 & -0.03$\pm$0.05 &  0.03$\pm$0.02 & -0.16 \\ % & 0.07\\
$\sigma_*(\Delta$[Zr/Fe])       & 0.046          & 0.01 & 0.09    & 0.09\\
$\sigma_*$(DS) & 0.055\\
$s_{\rm lines}(\Delta$[Zr/Fe])   & 0.06          & 0.07 & 0.06 & 0.11 \\
$\sigma_{*,{\rm ext}}$           & $<0.05$       & $<0.01$ & 0.07   &  $<0.09$ \\
\hline \\
$\langle\Delta$[Ba/Fe]$\rangle$ & 0.04$\pm$0.02 & $0.03\pm0.06$  & -0.01$\pm$0.05  & 0.03$\pm$0.03 & 0.12 \\ % & 0.38\\
$\sigma_*(\Delta$[Ba/Fe])       & 0.044         & 0.08 & 0.08   & 0.13 \\
$\sigma_*$(DS) & 0.049\\
$s_{\rm lines}(\Delta$[Ba/Fe])   & 0.08          & 0.07 & 0.11 & 0.10 \\
$\sigma_{*,{\rm ext}}$           & $<0.04$       & 0.04 & $<0.08$    & 0.08\\
\hline \\
$\langle\Delta$[La/Fe]$\rangle$ & 0.01$\pm$0.01 & 0.02$\pm$0.03 & -0.04$\pm$0.06  & 0.10$\pm$0.02 & -0.17 \\ % & 0.04 \\
$\sigma_*(\Delta$[La/Fe])       & 0.033          & 0.04 & 0.11    & 0.10 \\
$\sigma_*$(DS) & 0.025\\
$s_{\rm lines}(\Delta$[La/Fe])   & 0.06          & 0.08 & 0.07  & 0.07 \\
$\sigma_{*,{\rm ext}}$           & $<0.03$       & $<0.04$ & 0.08   & 0.07\\
\hline \\
$\langle\Delta$[Ce/Fe]$\rangle$ & 0.03$\pm$0.02 & 0.02$\pm$0.03 & -0.08$\pm$0.06 &  0.02$\pm$0.03 & -0.01 \\ %& 0.15\\
$\sigma_*(\Delta$[Ce/Fe])       & 0.041          & 0.04 & 0.10    & 0.13\\
$\sigma_*$(DS) & 0.025\\
$s_{\rm lines}(\Delta$[Ce/Fe])   & 0.10          & 0.02 & 0.07 & 0.08 \\
$\sigma_{*,{\rm ext}}$           & $<0.04$       & 0.04 & 0.08  & 0.10\\
\hline \\
\end{tabular} 
\end{table*}

\addtocounter{table}{-1}
\begin{table*}
\caption[]{Continued.
}
\begin{tabular}{lllllllll}
\hline
\hline
         & Cluster  & Evaporated   & Possible    &Stream/field & HD 26756 \% solar\\
        &           &    candidates     & cluster     & population &  (Asplund et al. 2005) \\
\hline
 $N$       & 6      & 2           & 3                  & 19\\  
 $F2$& $< 2$ &$< 2$& $\ge2$ &  $\ge 2$\\  
\hline
\multicolumn{2}{l}{Abundance}\\
\%HD~26756\medskip\\
\hline
$\langle\Delta$[Nd/Fe]$\rangle$ & 0.02$\pm$0.04 &0.01$\pm$0.01 & -0.05$\pm$0.09  & 0.12$\pm$0.04 & -0.16 \\ %& 0.05\\
$\sigma_*(\Delta$[Nd/Fe])       & 0.087          & 0.02 &  0.16    & 0.15 \\
$\sigma_*$(DS) & 0.032\\
$s_{\rm lines}(\Delta$[Nd/Fe])   & 0.05          & 0.07 & 0.07 & 0.07 \\
$\sigma_{*,{\rm ext}}$           & 0.07          &$<0.02$ &  0.15    & 0.14\\
\hline \\
$\langle\Delta$[Eu/Fe]$\rangle$ & $-0.04\pm$0.01& 0.07$\pm$0.01 & $-0.07\pm0.07$  & 0.05$\pm$0.02 & -0.10 \\ % & 0.08\\
$\sigma_*(\Delta$[Eu/Fe])       & 0.028          & 0.01 & 0.13   &  0.09 \\
$s_{\rm lines}(\Delta$[Eu/Fe])   & 0.06          & 0.06 & 0.06  &  0.06 \\
$\sigma_{*,{\rm ext}}$           & 0.03          &$<0.01$ &  0.11   &  0.07 \\ 
\hline  
\end{tabular} 
\end{table*}

\renewcommand{\baselinestretch}{1.0}

The main goal of these abundance measurements is to check whether at least a small fraction of the Hyades stream is compatible with having its origin in the Hyades cluster. To identify possible evaporated stars from the primordial cluster, we 
computed the $\chi^2$ statistics:
\begin{eqnarray}
\chi^2 (1) & = &\Sigma_{i={\rm Zr,Ba,La,Ce}}\;  \nonumber\\
     & & \left[ \left( [X_{i}/{\rm Fe}]\; - <[X_{i}/{\rm Fe}]>_{\rm Cluster}\right)/\sigma_{i,{\rm cluster*}} \right]^{2},
\end{eqnarray}
and 
\begin{eqnarray}
\chi^2 (2) & = &\Sigma_{i={\rm Zr,Ba,La,Ce}}\;  \nonumber\\
     & & \left[ \left( [X_{i}/{\rm Fe}]\; - <[X_{i}/{\rm Fe}]>_{\rm Cluster}\right)/\sigma_{i,{\rm cluster*}} \right]^{2}\\
     & & +  \left[ \left( [{\rm Fe I}/{\rm H}]\; - <[{\rm Fe I}/{\rm H}]>_{\rm Cluster} \right)/\sigma_{\rm FeI, cluster*} \right]^{2},\nonumber
\end{eqnarray}
where the average abundance for the cluster and the star-to-star scatter $\sigma_{\rm cluster*}$ are taken over the 5 sure members (see Table~\ref{Tab_param}). Despite the small number of stars involved, the differential abundance scatter used in the above $\chi^2$ formulae agree reasonably well with the values derived by Paulson et al. (2003) and De Silva et al. (2006) on a much larger sample of Hyades cluster stars [see $\sigma_*$(DS) and $\sigma_*$(P03) in Table~\ref{meanvalues_delta1}]. Thus the relatively small number of Hyades cluster stars used in the present study  does not appear to hamper our analysis.

The motivation to use two statistical indices [$\chi^2 (1)$ and $\chi^2 (2)$] based on different elemental sets is to illustrate the sensitivity 
to specific choices of elemental sets, as we will discuss below.
 We decided to leave Li, Nd and Eu out of these indices, because 
\begin{itemize}
\item  Nd possibly exhibits an abundance trend with $T_{\rm eff}$  (see Fig.~\ref{diff}) which could induce a spurious difference  between the average Nd abundances in the stream and cluster, depending on how stars in these two groups distribute with $T_{\rm eff}$;
\item Eu has the largest sensitivity to errors on the stellar-model parameters (Table~\ref{TabErr},  where $\Delta_{\rm err} = 0.15$~dex for Eu as compared to less than 0.10~dex for all the other elements);
\item Li is a very sensitive tagger which will be used afterwards (Sect.~\ref{Sect:Li}) as an independent check of the quality of the tagging based on the $\chi^2$ statistics from Fe and the heavier elements. 
\end{itemize}
The different number of degrees of freedom (4 and 5, where we suppose that the  average and $\sigma_*$ do not act to decrease the number of degrees of freedom; this is certainly true for the stream stars) for the two $\chi^2$ statistics makes their direct comparison difficult. It is therefore useful 
to use the $F2$ goodness-of-fit statistics instead, which transforms a $\chi^2$ statistics into a normal distribution with zero mean and unit variance, irrespective of the number of degrees of freedom. $F2$ is defined as 
\begin{equation}\label{Eq:F2}
F2 =  \left(\frac{9\nu}{2}\right)^{1/2} \left[\left(\frac{\chi^2}{\nu}\right)^{1/3}+\frac{2}{9\nu}-1\right],
\end{equation}
where $\nu$ is the number of degrees of freedom 
of the $\chi^2$ variable. The above definition corresponds to  the `cube-root transformation' of the $\chi^2$ variable (Stuart \& Ord 1994).

The values of $F2 (1)$ and $F2 (2)$ for all the stars studied are given in Table~\ref{deviations}. 
As apparent from that Table, all sure Hyades cluster members (plus the possible member HD~20430) have $F2 < 2$. This is actually not really surprising,  since the $\sigma_*$ value entering  the $\chi^2$ is derived from the cluster sample itself. The predictive power of this $\chi^2$ criterion therefore rather lies in its application to the stream sample. 
 Two stars from the stream (HD~149028 and HD~162808) have both $F2 < 2$ and may thus be considered as evaporated from the cluster (the analysis of the Li sequence presented in Sect.~\ref{Sect:Li} will confirm that conclusion).  All other stars, including the possible members (HIP~13600, HD~20439, and HD~26257), have  (at least one) $F2 > 2$ and must not be considered as cluster members based on the Fe and heavy element abundances, the latter star HD~26257 being even located way off the threshold, with $F2$ of the order of 7. This star was studied by De Silva et al. (2006) as well, who find relative abundances\footnote{The values quoted here are differences from their Table~2, with HD~26756 as the reference star.}  $\Delta$[Zr/Fe] = $-0.14$, $\Delta$[Ba/Fe] = $0.01$, $\Delta$[Ce/Fe] = $-0.20$, and $\Delta$[Nd/Fe] = $-0.05$ not so different form ours:  $\Delta$[Zr/Fe] = $-0.12$, $\Delta$[Ba/Fe] = $-0.12$, $\Delta$[Ce/Fe] = $-0.19$, and $\Delta$[Nd/Fe] = $-0.24$, the latter set being consistently underabundant, in contrast with the De Silva et al. values.
Actually, one may wonder whether  the three controversial Hyades cluster stars (HIP~13600, HD~20439,  and especially  HD~26257) could not in fact be  stream stars happening to be located in the spatial vicinity of the cluster. 
Interestingly, we stress that doubts were raised as well by De Silva et al. (2006) about the membership of HD~20439 (also known as vB~2), based on outlying abundances, in agreement with its outlying $F2$ obtained in the present study. Incidentally, these authors raise the same doubts for HD~20430 (vB~1), which our study does not tag as an outlier, however.

%Regarding the sensitivity of our conclusions to the selected elemental set, we note that adding Na or Mg  would not change any of the conclusions (and the %reason thereof will be apparent on Fig.~\ref{comp}), but adding Eu would make HD~149028 and HD~162808 disappear from the evaporated set, because %their Eu abundance is offset with respect to the cluster stars (see Table~\ref{TabAb} and Figs.~\ref{diff} and \ref{comp}). 

\begin{figure}
\includegraphics[width=7.5cm]{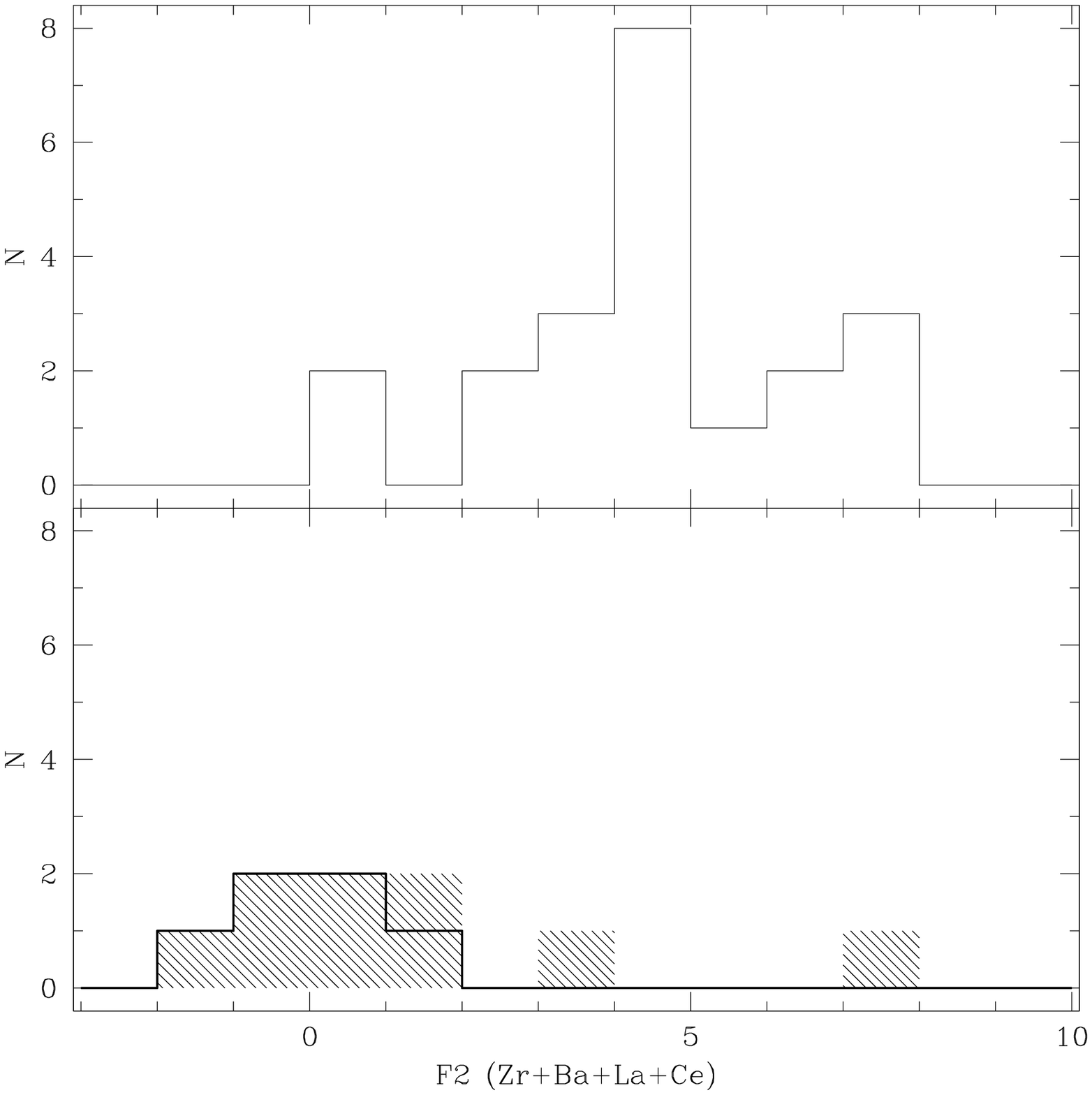}
\includegraphics[width=7.5cm]{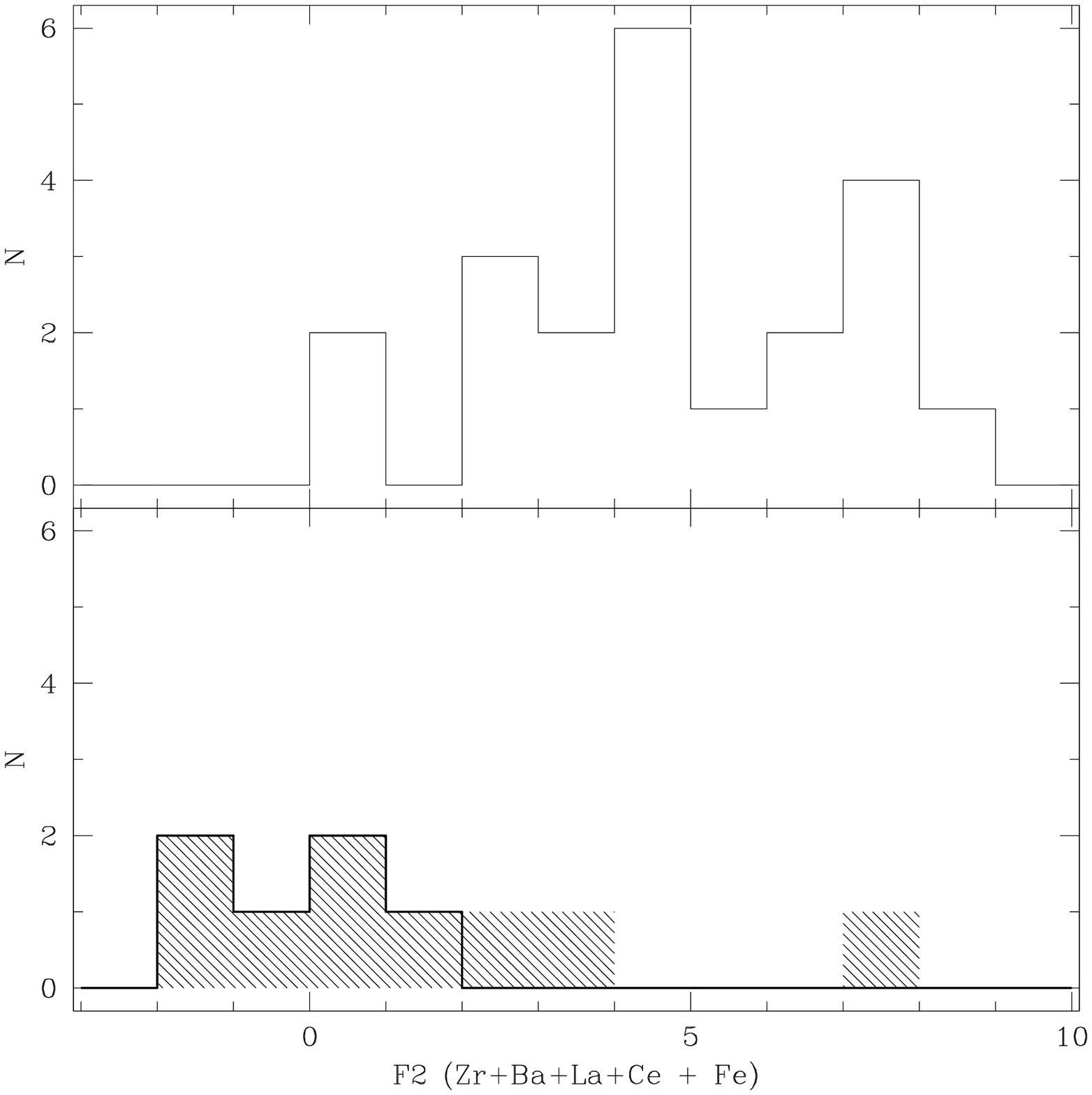}
\includegraphics[width=7.5cm]{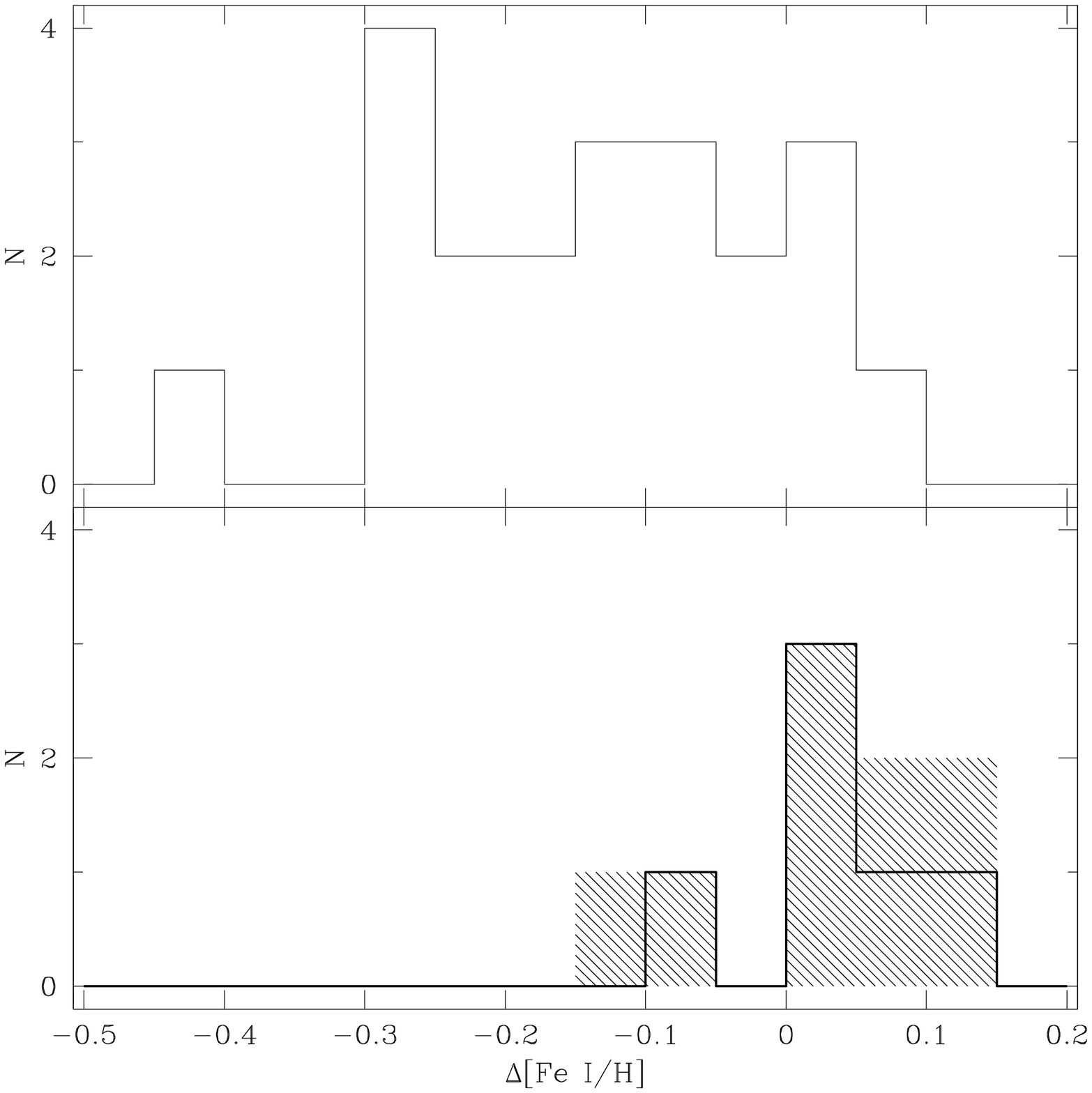}
\caption[]{\label{Fig:delta}
The distributions of $F2(1)$ (top panel), $F2(2)$ (middle panel) and $\Delta$~[Fe/H] (bottom panel) for the Hyades stream stars (open histogram),
cluster stars (shaded histogram), and sure cluster stars (thick histogram).
}
\end{figure}

In any case, there is a large population of stream stars which are chemically very different from the Hyades cluster members, as apparent from Fig.~\ref{Fig:delta} presenting the distributions of $F2 (1)$, $F2 (2)$ and $\Delta$ [Fe/H] for the Hyades stream and cluster. 
We may thus conclude that, irrespective of the index being used, only  a few stars from the Hyades stream have abundances compatible with the abundance ratios of stars of the Hyades cluster (see Sect.~8.1).

\begin{figure*}
\includegraphics[width=15cm]{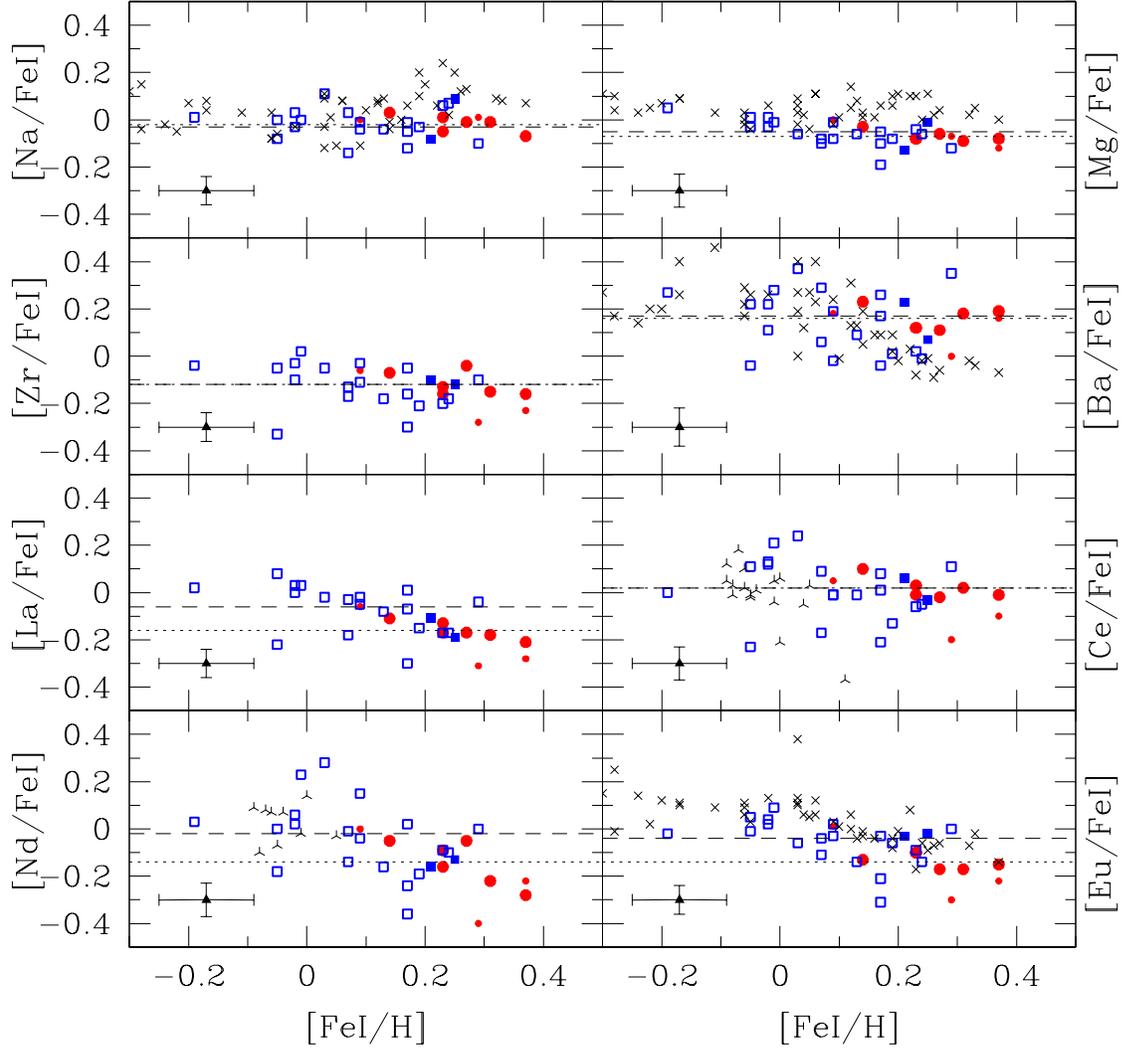}
\caption{[X/Fe] ratios vs. [Fe/H] for thin disk stars (4-branch crosses, from Bensby et al. 2003, 2005 and 3-branch crosses from Reddy et al. 2003 for Ce and Nd), certain Hyades cluster members (filled dots), possible Hyades members (open circles), 
evaporated candidates (filled squares), and stream/field stars  (open squares). All abundances in this work are relative to the Asplund et al. (2005) solar abundances (No attempt has been made to normalise the Bensby et al. and Reddy et al. data in the same way, since their adopted zero-points depend on the spectrograph -- see text. Slight offsets between the data sets may thus be expected). The dotted and dashed lines depict the average cluster and stream abundances, respectively (For Zr and Ce, they overlap). The error bars drawn in each panel represent 
the abundance uncertainties due to the line-to-line scatter.
}
\label{comp}
\end{figure*}

\section{Lithium as an efficient population tagger}
\label{Sect:Li}

\begin{figure}
\includegraphics[width=8cm]{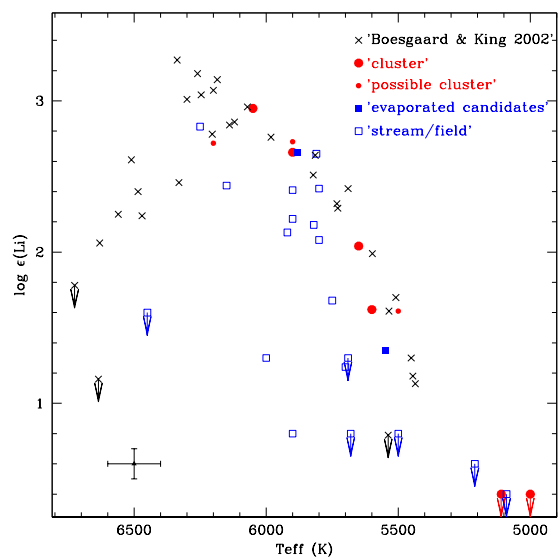}
\caption{The Li abundance for the different stellar categories as flagged from the $F2$ indices, compared to Boesgaard \& King (2002) Li abundances for Hyades cluster stars (crosses). Arrows mark upper limits on the Li abundance.
 \label{Fig:Li}
}
\end{figure}

We now turn to lithium  as a tagger that we use  to check the quality of the tagging based on Fe and heavier elements. Boesgaard \& King (2002) have provided a comparison sample of Li abundances in Hyades cluster stars, as shown on Fig.~\ref{Fig:Li} (crosses). The Li abundance in the cluster stars follows a tight sequence that goes through a maximum around   6200~K, and decreases sharply on either side as Li gets destroyed by diffusive or convective downward motions (see Deliyannis 2000 and Sestito \& Randich 2005 for reviews). On the opposite, in field stars, the Li abundance spans a substantial range (of the order of 2~dex) for any given effective temperature ({\it e.g.,} Israelian et al. 2009 and references therein), especially on the cooler side of the peak.
The tight sequence observed for the Hyades cluster is a direct consequence of the fact that its stars are coeval (Deliyannis 2000). It offers an efficient way to tag stream  stars evaporated from the cluster (although one cannot totally exclude the possibility that a stream star falls by chance along the Hyades sequence).

Fig.~\ref{Fig:Li} confirms the chemical tagging analysis performed in Sect.~\ref{Sect:evap}, since  the two stars tagged as `evaporated candidates' indeed lie along the Hyades cluster Li sequence. Conversely, we find three stream stars falling along the Li sequence (HD~25680, HD~142072, HD~149285), but this can be a chance occurrence, and does not impose a revision of our former tagging (among these three, only HD~25680 has both $F2 < 4$). Finally, we note that the three  `possible cluster members' nicely fall along the Hyades Li sequence.

\section{Discussion}

\subsection{Can the stream fully originate from the cluster?}

Since Famaey et al.~(2007) estimated that the Hyades overdensity in velocity space constitutes $p_0= 75$\% of the Hyades box (this frequency being almost independent of the colour range), we can ask ourselves whether this overdensity can {\it entirely} be due to the population of stars evaporated from the Hyades cluster. 

However, if this were true, we would expect to have found about 15 (or about 11 to 19 with Poissonian uncertainties) stars with abundances compatible with the Hyades cluster ($p_0 N=15.75$ with $N=21$), instead of 2, in the most favorable case. This difference leads to a significance level close to $\alpha = 1.00$ (within $10^{-9}$) for the unilateral rejection of the null hypothesis 
that all the stars of the velocity overdensity are originating from the Hyades cluster, as obtained from the relation  (resulting from the fact that for $p_0= 0.75$ and $N = 21$, the binomial distribution is already well approximated by a Gaussian distribution)
\begin{equation}
u_{\rm obs} = \frac{|x - N p_0|}{(N p_0(1-p_0))^{1/2}}      \ge u_{\alpha}, 
\end{equation} 
where $x=2$, $u_{\rm obs} = 6.9$ and $u_{\alpha}$ is the abscissa of the reduced 
normal distribution ${\cal N}(0,1)$ such that $\int_{-\infty}^{u_{\alpha}}{\cal N}\mathrm(0,1) = \alpha$. 

We are confident that the sample selection criterion $V_{\rm 
rot} \sin i \le 10 \kms$ (Sect.~\ref{Sect:sample}) does not affect this conclusion, for the following reason:
all stream stars of our sample are low-mass main sequence stars with  $b-y \ge 0.3$ (or equivalently  $B-V \ge 0.5$, according to Tables~15.7 and 15.10 of Drilling \& Landolt 1999),  
and Paulson et al. (2003) showed that such stars in the Hyades cluster indeed have 
$V_{\rm rot} \sin i \le 10 \kms$, so that no candidate evaporated star  should have been missed in the stream by rejecting the fast rotators. 

This is thus an independent confirmation of the dynamical (most probably resonant) origin of most of the Hyades stream, even though a population of stars evaporated from the Hyades cluster does overlap with this dynamical stream.

\subsection{Properties of the distinct populations of the stream}
\label{Sect:stream}

Fig.~\ref{diff} presents the 
differential abundances $\Delta$[X/Fe] (relative to HD~26756) against
$T_{\rm eff}$ with different symbols identifying the
Hyades cluster members (large filled circles), the possible cluster members
(small filled circles), evaporated candidates (large filled
squares), and
finally stars from the stream/field (open squares). 
In Table~\ref{meanvalues_delta1}, we
evaluate the average abundances for each of the elements considered,  
separately for these four groups.
% the Hyades cluster (sure members +
%possible members), the evaporated
%stars (Table~\ref{meanvalues_delta1} is based on the $\delta_1$ membership
%criterion, and Table~\ref{meanvalues_delta2} on $\delta_2$; see
%Sect.~6 for details), and the field/stream stars. 
We report the mean and standard
deviations $\sigma_\star$ for a given chemical element separately for
these four samples, and the root-mean-square $s_{\rm lines}$ (taken
over all stars of a given group) of the line-to-line scatter $\sigma$
(for a given star, taken from Tables~\ref{Tab_param} and
\ref{TabAb}). The $s_{\rm lines}$ parameter must be seen as an
instrumental scatter, to which any extrinsic (physical) scatter
($\sigma_{*,{\rm ext}}$) will
add quadratically to yield $\sigma_\star$, so that we must expect $\sigma_\star \ge s_{\rm
  lines}$. For the `cluster' and `evaporated candidates' samples however, $\sigma_{*}$ is usually smaller than $s_{\rm
  lines}$, which  indicates that the true scatter of these samples is so small that it cannot be derived accurately with the achieved line-to-line scatter.   

A very consistent picture thus emerges from
Table~\ref{meanvalues_delta1} regarding the sample genuine scatter
$\sigma_{*,{\rm ext}}$: for the cluster sample, it is, as expected, quite small ($\le 0.05$~dex) for all elements
but Fe~I and Nd, where 
it reaches 0.07~dex. Except for these two elements, our results agree well with the conclusion of chemical homogeneity found for 46 Hyades cluster stars by Paulson et al. (2003) and De Silva et al. (2006; compare the entries listed as $\sigma_*$ in Table~\ref{meanvalues_delta1} with $\sigma_*$(DS) or $\sigma_*$(P03)).
On the contrary, for the stream,  the scatter is systematically much larger, ranging from 0.05~dex for Mg to 0.15~dex for Nd, as expected if the stream is dominated by {\it field} stars.
It is not surprising either that the sample of evaporated stars has properties similar to those of the cluster, both in terms of scatter and average value, since they were actually selected to ensure such a similarity. The `Possible cluster' stars deviate markedly from the cluster values, but as indicated in Sect.~\ref{Sect:evap}, the statistical properties of this group are dominated by the outlying star HD~26257.
There are as well noteworthy differences (by a least 0.1~dex) between cluster and stream stars in terms of their average abundances, especially for Fe, La, Nd and Eu (Table~\ref{meanvalues_delta1}).  The difference in the average metallicity of the two groups (0.2~dex) is very significant (Table~\ref{meanvalues_delta1} and Fig.~\ref{Fig:delta}). As apparent from Fig.~\ref{comp}, 
the difference in the average La, Nd, and Eu abundances between the Hyades cluster and stream appears as a natural consequence of their different average metallicities and of the trend with metallicity observed for these elements in the thin disk. 
%Similarly, the large scatter observed for Ba appears  to be a natural consequence of the steeper slope of the Ba -- metallicity trend.
This finding is an important result of the present paper:
the stream/field population (thus cleaned from the evaporated metal-rich cluster stars)   
remains more metal-rich
([Fe/H] of the order 0.04 to 0.08, normalized to a solar Fe abundance of 7.45, or -0.01 to 0.05 with the more traditional normalization of 7.50 from Grevesse \& Sauval (2000), in better agreement with older studies; Table~\ref{meanvalues_delta1}) than the  
thin disk population in the solar neighbourhood. 
According to Famaey et al. (2007), the mean (photometric) metallicity of the whole Geneva-Copenhagen survey 
for the solar neighbourhood is [Fe/H]~$= -0.16$. Excluding halo stars 
with [Fe/H] $< -0.5$, and excluding all the stars from the Hyades box, 
we get a mean (photometric) metallicity of [Fe/H]~$ = -0.12$. Given that photometric metallicity estimates 
are systematically smaller than spectroscopic values by about 0.05~dex (Holmberg et al. 2009), the thin disk metallicity in the solar neighbourhood may be evaluated at $-0.07$~dex  (see also Allende-Prieto 2010 and references
therein). 
Comparing this thin-disk metallicity of  $-0.07$~dex with that of the stream/field
(of the order $-0.01$ to 0.08~dex) clearly supports an inner-disk
origin for the stream. We stress that this comparison is somewhat biased by the fact that about 25\% of the stars in the Hyades velocity box must belong to the local background Schwarzschild velocity ellipsoid (Sect.~1 and Famaey et al. 2007), and thus have a metallicity typical of the thin disk. One may try to clean the sample from the local background field stars to better evaluate the metallicity of the stream stars. An upper limit to the average stream metallicity may be obtained by removing from the 21 stars making up our Hyades box   the 25\%  most metal-deficient stars, believed to belong to the field. Thus removing 
the most metal-poor stars HD~86165, HD~103891, HD~133430, HD~149285 and HD~171067 from the stream/field sample, one finds an average metallicity of 
0.114~dex (with the $\log \epsilon = 7.45$ normalization), corresponding to an excess of (at most) 0.18~dex with respect to the metallicity of the local thin disk. 

The argument of an inner-disk
origin for the stream  may even be made more quantitative 
by adopting a reasonable metallicity gradient of $-0.05$ to
$-0.07$~dex/kpc in the galactic disk (e.g. Daflon \& Cunha 2004): this
would indicate that the Hyades stream originates at $\sim$0.9 to 3.0~kpc (or even at 3.6~kpc if one adopts the upper limit to the metallicity excess,  or even more taking into account the fact that the gradient for old stars may be even flatter)
inwards in the Galactic disk. In Sect.~\ref{Sect:ILR}, this range will be compared with predictions made in the framework of a stream 
caused by an inner 4:1 resonance from a 2-armed spiral pattern (inner ultra-harmonic resonance or IUHR). 

The large difference in the average metallicities of the Hyades cluster and stream
(after cleaning the latter from the 2 evaporated candidates)  definitely
and undoubtedly confirms that the stream is not solely made
out of evaporated cluster stars. And concomittantly, this finding
reinforces the hypothesis of the dynamical nature of the Hyades
stream. In this context, one may thus compare the scatter and abundance trends obtained for the Hyades stream stars with those in the
inner thin disk from which they originate. This is done in Fig.~\ref{comp} using the samples of Bensby et al. (2003, 2005) and Reddy et al. (2003). 
It is seen that the stream conforms with the properties of the thin disk: absence of metallicity trend and small scatter for Mg, relatively large scatter for Ba, and small scatter and well marked trend for Eu (Fig.~\ref{comp}). The small shifts between our values and Bensby's ones, apparent for Mg and Eu, may be attributed to the slightly different line list and solar abundances adopted for normalization, with $\log \epsilon$ of 7.53, 7.45 and 0.52 adopted in this work for Mg, Fe, and Eu respectively, as compared to 7.57 -- 7.58, 7.53 -- 7.56, and 0.47 -- 0.56 for Bensby et al. (2005; their Table~5, where the adopted abundances may depend upon the spectrograph and line used).

\subsection{A resonance from the spiral pattern?}
\label{Sect:ILR}

\begin{figure}
\centering
\includegraphics[width=8cm]{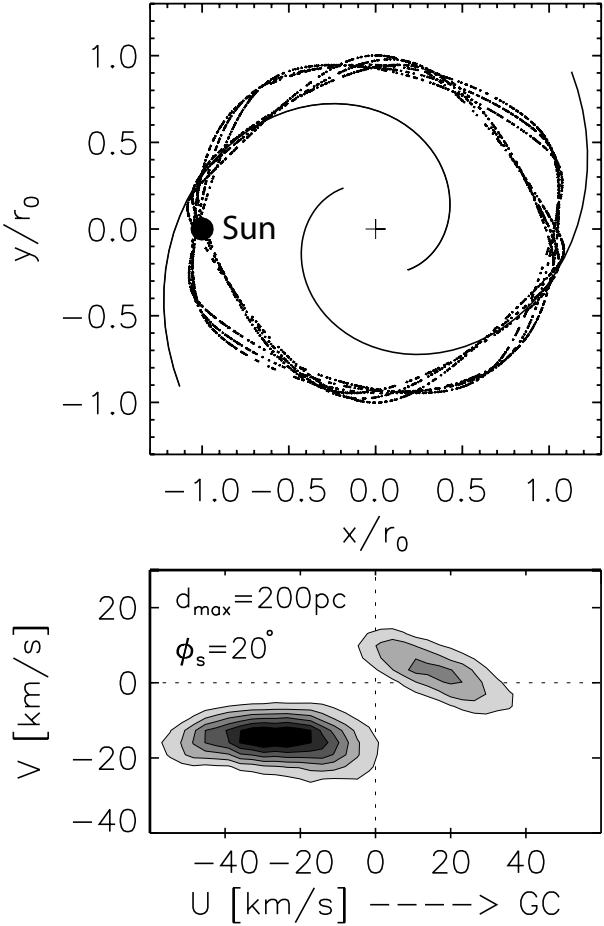}
\caption{Top panel -- The effect of a 2-armed spiral structure on
orbits near the 4:1 IUHR. Note the splitting into 2 families of closed
orbits in the
frame moving with the (trailing) spiral pattern. For a Sun orientation at $20^\circ$
with respect to a
concave arm, both orbital families enter the solar neighborhood stellar velocity
distribution (black filled circle). The galactocentric axes are in units of $r_0$ (the galactocentric radius of the Sun). Bottom panel -- The effect on the
$UV$-plane for the configuration shown in the top panel (selecting test particles in a 200~pc circle around the Sun). Each orbital
family gives rise to a stream in velocity space. We can associate the dense clump at $(U,V)\approx(-35,-17)$ km/s with the Hyades and the shallow one  at $(U,V)\approx(10,0)$ km/s with Sirius. The contour levels
correspond to 0.2, 0.31, 0.43, 0.55, 0.67, and 0.8 of the maximum value at the centre of the Hyades clump.
 }
\label{fig:hyades_uv}
\end{figure}

The question now is which dynamical effects in the Galaxy can both place the Hyades stream in its observed location in velocity space, while also having a metallicity typical of stars originating at 0.9 to 3.0~kpc inside the solar circle. 

As was shown by Minchev et al.~(2010), a resonance with
the Galactic bar cannot account for the position of the Hyades stream in the $UV$-plane (while it actually can account for the Pleiades), nor
for an origin farther away than $\sim0.8$~kpc in the inner disk. On the other hand, using an orbital weighting function technique, Quillen \& Minchev~(2005) showed that the 4:1 inner ultra-harmonic resonance (IUHR) of a 2-armed spiral structure\footnote{Observations indicate that the Milky Way has a 4-armed structure, but with 2 more prominent arms (see, e.g., Englmaier et al. 2008).} splits the velocity distribution
into two features
corresponding to two orbital families, one of them consistent with the
Hyades stream (see also Sect.~\ref{sec:intro})\footnote{Note that this splitting into different orbital families of different average galactic radii is also seen at the 4:1 inner Lindblad resonance of a 4-armed pattern (see, e.g., Amaral \& L\'epine 1997).}. To check this possibility, we performed test-particle simulations of a stellar disk consistent with the Milky Way kinematics, perturbed by a 2-armed spiral pattern. Details
about the simulation parameters can be found in Minchev \& Quillen (2007).
We examined a range of pattern speeds $\Omega_s$, and indeed found that we could reproduce the position of the
Hyades stream in the $UV$-plane (using the solar motion of Sch\"onrich et al. 2010) only when the solar circle is near the 4:1~IUHR. Fig.~\ref{fig:hyades_uv} presents our results for
$\Omega_s=18$ km/s/kpc for an angular speed at the solar position $\Omega_0=28$ km/s/kpc (i.e. $\Omega_s/\Omega_0=0.65$).
The top panel shows the two orbital families and the bottom panel the
resulting velocity
distribution in the $UV$-plane. 

The beauty of this simulation is that, while reproducing the Hyades stream, the other orbital family creates another remarkable feature in velocity space around $(U,V)\approx(10,0)$~km/s, which is consistent with the Sirius moving group (see e.g. Famaey et al. 2005, 2008). It is slightly shallower than the Hyades stream, exactly as observed. 
This simulation furthermore predicts that the Hyades stream originates about 1~kpc inward from the solar radius, which lies within the error bar 
derived in Sect.~\ref{Sect:stream} from the observed metallicity difference (local thin disk - Hyades stream: 0.06 to 0.15~dex) and metallicity gradient in the thin disk ($-0.05$ to $-0.07$~dex/kpc; Daflon \& Cunha 2004). This error bar is unfortunately quite large still (0.9~--~3.0~kpc) to meaningfully constrain  the simulation, and one should strive to reduce it in the future.
If in the end, a metallicity difference as large as (or larger than) 0.1~dex were confirmed, agreement with the present simulation could only be restored at the expense of the necessity for a steeper metallicity gradient in the disk. 
Let us however note that Daflon \& Cunha (2004) studied OB stars, i.e. 
the very recent metallicity gradient, and that the gradient for older 
stars such as those of the Hyades stream is expected to be flatter due 
to radial mixing, not steeper. This might cast doubt on the ability of 
the present dynamical model to reproduce the observed metallicity 
excess, should it be confirmed by further studies. 
On the other hand, there exists other mechanisms that could have made the Hyades stream migrate from larger distances in the inner disk. It is known that
stars can exchange angular momentum in galactic disks due to three
distinct mechanisms:
transient spirals (Sellwood \& Binney~2002), spiral-bar resonance overlap
(Minchev \& Famaey~2010), and the effect of minor mergers on the disk
(Quillen et al.~2009).
One of these could be responsible for migrating the Hyades stream from
a distance larger than
what is possible with a spiral structure model only. The ``migrating'' stream could remain coherent for a long enough time before dispersing, but more work is needed to investigate this possibility. The drawback would then be that the Hyades and Sirius overdensities in the $UV$-plane would be unrelated and would need two separate explanations.

\section{Conclusion}

From an analysis of the Li, Na, Mg, Zr, Ba, La, Ce, Nd and Eu abundances of 21 stars from the Hyades stream, we performed chemical tagging to identify stream stars possibly evaporated from the cluster. The first tagging method is based on a $\chi^2$ test  comparing the abundances of  Fe, Zr, Ba, La and Ce in stream stars with the cluster average value and internal scatter. It is convenient to make use of the `cube-root transformation' of the $\chi^2$ variable to convert it into the goodness-of-fit statistics $F2$, which behaves as a normal distribution with zero mean and unit variance.  This method relies of course on an accurate evaluation of the  cluster average value and internal scatter. This tagging method  works best for chemical elements exhibiting a steep slope with metallicity. Any metallicity difference between the stream and the cluster is then amplified by each element involved in the $\chi^2$ sum and exhibiting a steep metallicity trend. The absence of any such metallicity trend among Na and Mg (at least for metallicities in the range -0.1 -- 0.3) makes them useless for chemical tagging.   A second method uses the tight Li --$T_{\rm eff}$ sequence  observed among Hyades stars, a result of the slow Li destruction process at work in these stars. Stars of the same $T_{\rm eff}$  but of a different age, as will be the case in general for stream stars, will not fall along the Hyades sequence. This method offers the advantage that the range spanned by Li abundances is very large (of the order of 3~dex), but is restricted to stars in the narrow $T_{\rm eff}$ range 5000 -- 6500~K. Outside this range, the Li abundance is either too small to be measurable or not sensitive to age any longer.   Based on these methods, only two stars from the stream  (HD~149028 and HD~162808) appear to have abundances within `$2\sigma$' of the cluster (i.e., $F2 < 2$, and matching the Li sequence); they thus very likely evaporated from the cluster. 

Our analysis thus convincingly demonstrates that a large fraction of stars in the Hyades velocity box ($\sim 90$\%) cannot originate from the Hyades cluster. Interestingly, 2 evaporated candidates have been found, grossly in line with the conclusion of Famaey et al. (2007) that about 15\% of the Hyades stream could still originate in the Hyades cluster. It is then of high interest to clean the stream population from this evaporated population and to compare its properties with the local thin disk. It is found here that these stream stars are more metal-rich than the local thin disk, with an excess of the order of  $0.06$ to 0.15~dex. This metallicity excess implies an  origin for the stream at about 0.9~--~3.0~kpc inwards from the Sun, adopting a galactic metallicity gradient of $-0.05$ to $-0.07$~dex/kpc. Predictions from a 4:1 inner resonance  of a 2-armed spiral pattern  locate the origin of the Hyades stream at a  maximum of 1~kpc inwards from the solar radius. A more precise determination of the stream metallicity is thus needed to confirm or contradict this scenario. Besides reproducing the Hyades overdensity in the $U - V$ plane, this scenario however makes another appealing prediction, namely the existence of the Sirius moving group exactly as observed. 

As a conclusion, the analysis of this small subsample of 21 stars from the Hyades stream has already yielded some interesting insights on its nature, and it will be of prime importance to confirm these with much larger samples, e.g. in future analyses performed with multi-fibre spectrographs such as the High Resolution Multi-Object Spectrograph of the Anglo-Australian Telescope (Barden et al. 2008). If they could derive a more accurate metallicity for  the stream cleaned from the evaporated population, these large surveys could clearly provide a  meaningful constraint on the distance at which the Hyades stream originates, and on the metallicity gradient in the disk, thus shedding light on the dynamical process responsible for the formation of the stream. This, together with the mapping of large-scale non-axisymmetric motions (see, e.g., Siebert et al. 2011) will help better constraining the non-axisymmetry of the Galactic potential and its influence on the dynamical and chemical evolution of the disk.

\section*{acknowledgements}
The HERMES/Mercator project is a collaboration between the K.U. Leuven, the Universit\'e libre de Bruxelles and the Royal Observatory of Belgium with contributions from the Observatoire de Gen\`eve (Switzerland) and the Th\"uringer Landessternwarte Tautenburg (Germany). It is funded by the Fund for Scientific Research of Flanders (FWO) under the grant G.0472.04, from the Research Council of K.U. Leuven under grant GST-B4443, from the Fonds 
National de la Recherche Scientifique (F.R.S.-FNRS) under contracts IISN4.4506.05 and FRFC 2.4533.09, and financial support from Lotto (2004) assigned to the Royal Observatory of Belgium. L.P. acknowledges financial support from the  \textit{D\'epartement des Relations Internationales} of the Universit\'e libre de Bruxelles
and thanks the F.R.S-FNRS (Belgium) for granting her a {\it Bourse de s\'ejour scientifique}. Support has also been provided by the {\it Actions de recherche concert\'ees (ARC)} (Heavy elements in the universe: 
stellar evolution, nucleosynthesis and abundance determinations)
from the \textit{Direction g\'en\'erale de l'Enseignement non obligatoire et de la Recherche scientifique - Direction de la Recherche scientifique - Communaut\'e fran\c{c}aise de Belgique}. B.F. acknowledges support from the CNRS (France) and AvH foundation (Germany). G.G. is a postdoctoral researcher of the FWO-Vlaanderen (Belgium). S. vE. is FNRS research associate. Authors contributions: LP, TM, SVE, AJ and CSn have conducted the 
spectroscopic analysis; BF, AJ, IM, AS, JRDL and OB have conducted the dynamical and statistical analyses; CSi, GG, TD and EP have conducted the observations; AJ, HVW, CW, GR, SP, WP, HH, YF, and LD are builders of the HERMES/Mercator spectrograph.

\bsp
\label{lastpage}
\end{document}